\pdfoutput=1
\documentclass[aps,twocolumn, floatfix, superscriptaddress]{revtex4}
\usepackage{graphicx}
\DeclareGraphicsExtensions{.pdf}
\usepackage{amsmath,amssymb,bbold,bm,color}
\usepackage{float}
\usepackage{epstopdf}
\usepackage{tabularx}
\usepackage{braket}

\begin{document}

\title{Quantum oscillations and Dirac-Landau levels in Weyl superconductors}

\author{Tianyu Liu}
\affiliation{Department of Physics and Astronomy, University of British Columbia, Vancouver, BC, Canada V6T 1Z1}
\affiliation{Quantum Matter Institute, University of British Columbia, Vancouver BC, Canada V6T 1Z4}
\affiliation{Department of Materials Engineering Science, Osaka University, Toyonaka 560-8531, Japan}
\author{M. Franz}
\affiliation{Department of Physics and Astronomy, University of British Columbia, Vancouver, BC, Canada V6T 1Z1}
\affiliation{Quantum Matter Institute, University of British Columbia, Vancouver BC, Canada V6T 1Z4}
\author{Satoshi Fujimoto}
\affiliation{Department of Materials Engineering Science, Osaka University, Toyonaka 560-8531, Japan}

\begin{abstract} 
  When magnetic field is applied to metals and semimetals quantum oscillations appear as individual Landau levels cross the Fermi level. Quantum oscillations generally do not occur in superconductors (SC) because magnetic field is either expelled from the sample interior or, if strong enough, drives the material into the normal state. In addition, elementary excitations of a superconductor -- Bogoliubov quasiparticles -- do not carry a well defined electric charge and therefore do not couple in a simple way to the applied magnetic field.  We predict here that in Weyl superconductors certain types of elastic strain have the ability to induce chiral pseudo-magnetic field which can reorganize the electronic states into Dirac-Landau levels with  linear band crossings at low energy. The resulting quantum oscillations in the quasiparticle density of states and thermal conductivity can be experimentally observed under a bending deformation of a thin film Weyl SC and provide new insights into this fascinating family of materials.
\end{abstract}

\date{\today}

\maketitle

\section{Introduction}
Quantum oscillations \cite{shoenberg1984} furnish an essential experimental tool for measuring the Fermi surface of metals. They also help to understand electronic structures of the recently discovered topological insulators \cite{checkelsky2009, taskin2009, eto2010, analytis2010, qu2010} and topological Dirac and Weyl semimetals \cite{potter2013, he2014, moll2016, xiong2016}. However, probing superconductors by the quantum oscillation technique has been thought impossible because such measurements require strong magnetic fields which are either expelled from the SC due to the Meissner effect or render the material normal. Type-II superconductors  allow the field to penetrate but form the Abrikosov vortex state, whose quasiparticle eigenstates are known to be Bloch waves and not Landau levels \cite{franz2000,vafek2001, liu2015}. 

Quantum oscillations in resistivity \cite{barivsic2013, doiron2007, ramshaw2011}, Hall coefficient \cite{leboeuf2007}, thermal conductivity \cite{grissonnanche2014}, and torque \cite{jaudet2008} have already been observed in underdoped cuprates when magnetic field suppresses superconductivity. Quantum oscillations with $1/\sqrt{B}$ periodicity have also been predicted to appear in vortex lattice \cite{alexandrov2008} and vortex liquid states \cite{banerjee2013} in cuprates and are observed in 2H-NbSe$_2$ \cite{corcoran1994}. However, reports on conventional quantum oscillations periodic in $1/B$ in the superconducting state are lacking presumably due to the reasons listed above.  

We argue here that this difficulty can be overcome by using the recently proposed Dirac and Weyl superconductors \cite{yang2014, meng2012}, which possess unusual electronic structures comprising linearly dispersing quasiparticle bands at low energies, similar to graphene and  $d$-wave SC in two dimensions and to Dirac and Weyl semimetals in three dimensions. One may expect that  Dirac and Weyl superconductors will exhibit a variety of exotic behaviors similar to their semimetal counterparts and to $d$-wave SC. Specifically, it has long been known that elastic strain can induce chiral pseudo-magnetic field and Landau quantization in graphene \cite{guinea2010,levy2010}. Similar effects have been predicted to occur in 3D Dirac and Weyl semimetals \cite{cortijo2015, sumiyoshi2016, pikulin2016, grushin2016, cortijo2016, arjona2017,liu2017}, and very recently also in $d$-wave SC \cite{massarelli1,nica1}.

In this work, through a combination of analytical calculations and numerical simulations, we demonstrate that  quantum oscillations can also occur in Dirac and Weyl superconductors under certain types of elastic deformations at zero magnetic field.  Remarkably, these quantum oscillations arise due to the formation of Landau levels comprised of {\em charge neutral} Bogoliubov quasiparticles deep in the superconducting state. To support these findings we organize the paper as follows. In Section \ref{Model Hamiltonian}, we formulate a model of a Weyl superconductor and discuss its spectrum and phase diagram. In Section \ref{Strain induced gauge field}, we incorporate strain to our Hamiltonian and show that to leading order it produces  pseudo-magnetic field in the low-energy sector. In Section \ref{Longitudinal thermal conductivity}, we show that the strain-induced pseudo-magnetic field can give rise to quantum oscillations in density of states (DOS) and longitudinal thermal conductivity. Section \ref{Conclusion} concludes the paper by discussing the experimental feasibility in candidate materials and outlines various potentially interesting directions based on our current work.

\section{Model Hamiltonian} 
\label{Model Hamiltonian}
\begin{figure}
\includegraphics[width = 8.0cm]{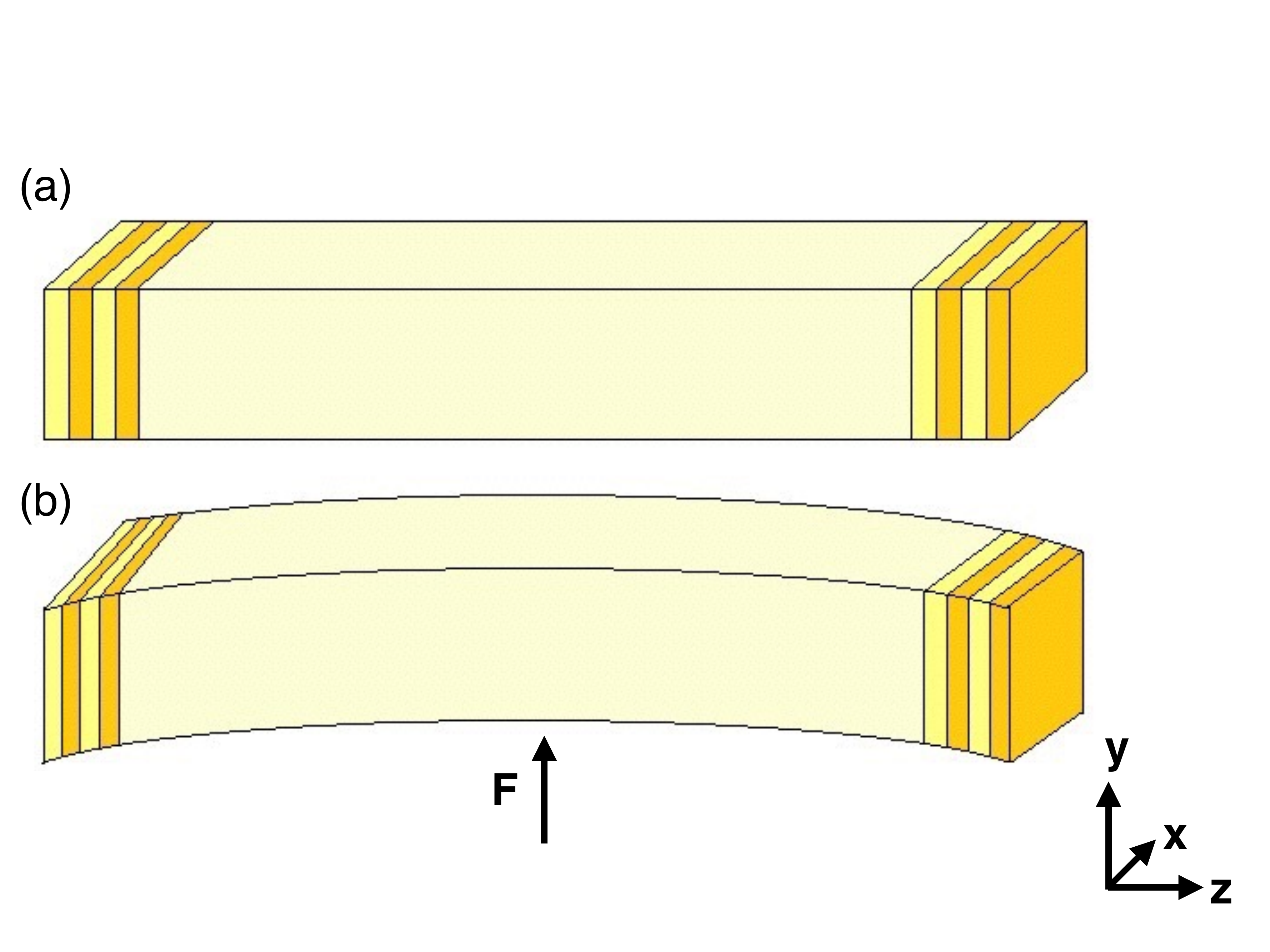}
\caption{Schematic plot for (a) undeformed and (b) bent TI-SC multilayer Weyl superconductor. The alternating TI and SC layers are omitted in the bulk but explicitly drawn at ends to illustrate that there are integer number of unit cells comprised of one TI layer and one SC layer.} \label{fig1}
\end{figure}

We employ the multilayer model of Meng and Balents \cite{meng2012} as illustrated in Fig.~\ref{fig1}. The model comprises alternating topological insulator (TI)  and $s$-wave SC layers stacked along the $z$-direction. For the TI layers, for simplicity, only the surface states are considered. In the following we modify the Meng-Balents model slightly by adding anisotropy to the  Zeeman mass term, which will allow us regularize the Hamiltonian on the tight binding lattice without adding extra Weyl points near the corners of the Brillouin zone.

The Hamiltonian of such a TI-SC multilayer system reads
\begin{equation} \label{H}
 H=H_{\text{TI}}+H_{\text{SC}}+H_{t_d}+H_{t_s}
\end{equation}
where
\begin{gather*}
H_{\text{TI}} = \sum\limits_{\bm k_\perp, z} \psi_{\bm k_\perp z}^\dagger[\hbar v_F \sigma^z (\hat z \times \bm s)\cdot \bm k_\perp + (m-m'a^2 k_\perp^2)s^z]\psi_{\bm k_\perp z} 
\\
H_{\text{SC}} =\sum\limits_{\bm k_\perp, z} (\Delta c_{\bm k_\perp  z, 1\uparrow}^\dagger c_{-\bm k_\perp  z, 1\downarrow}^\dagger+\Delta c_{\bm k_\perp  z, 2\uparrow}^\dagger c_{-\bm k_\perp  z, 2\downarrow}^\dagger )+{\rm h.c.}
\\
H_{t_d} =\sum\limits_{\bm k_\perp, z} \Big(\frac{1}{2}t_d \psi_{\bm k_\perp z+1}^\dagger \sigma^+ \psi_{\bm k_\perp z} +\frac{1}{2}t_d \psi_{\bm k_\perp z-1}^\dagger \sigma^- \psi_{\bm k_\perp z}\Big) 
\\
H_{t_s} =\sum\limits_{\bm k_\perp, z} \psi_{\bm k_\perp z}^\dagger t_s \sigma^x \psi_{\bm k_\perp z} 
\end{gather*}
The basis $\psi_{\bm k_\perp z}=(c_{\bm k_\perp  z, 1\uparrow}, c_{\bm k_\perp  z, 1\downarrow}, c_{\bm k_\perp  z, 2\uparrow},c_{\bm k_\perp  z, 2\downarrow})^T$ is written in terms of annihilation operators $c_{\bm k_\perp  z, \sigma s}$ for electrons located in the $z$-th unit cell with an in-plane momentum $\bm k_\perp = (k_x, k_y)$ and spin projection $s_z=\uparrow,\downarrow$. ``Sublattice'' labels $\sigma_z=1,2$ specify the TI-SC interface in the single unit cell. Pauli matrices $\bm s$ and $\bm \sigma$ act in spin and sublattice space, respectively. Physically, $H_{\text{TI}}$, $H_{\text{SC}}$, $H_{t_d}$, and $H_{t_s}$ can be interpreted as describing the Zeeman gapped topological insulator surface states, proximity-induced pairing, hopping between adjacent unit cells, and hopping within a single unit cell, respectively. The $m'$ term in $H_{\text{TI}}$ represents the above mentioned modification of the Meng-Balents model (it is easy to check that it has no significant effect at small $k$ as long as $m'$ is chosen appropriately small).

\begin{figure}
\includegraphics[width = 8.0cm]{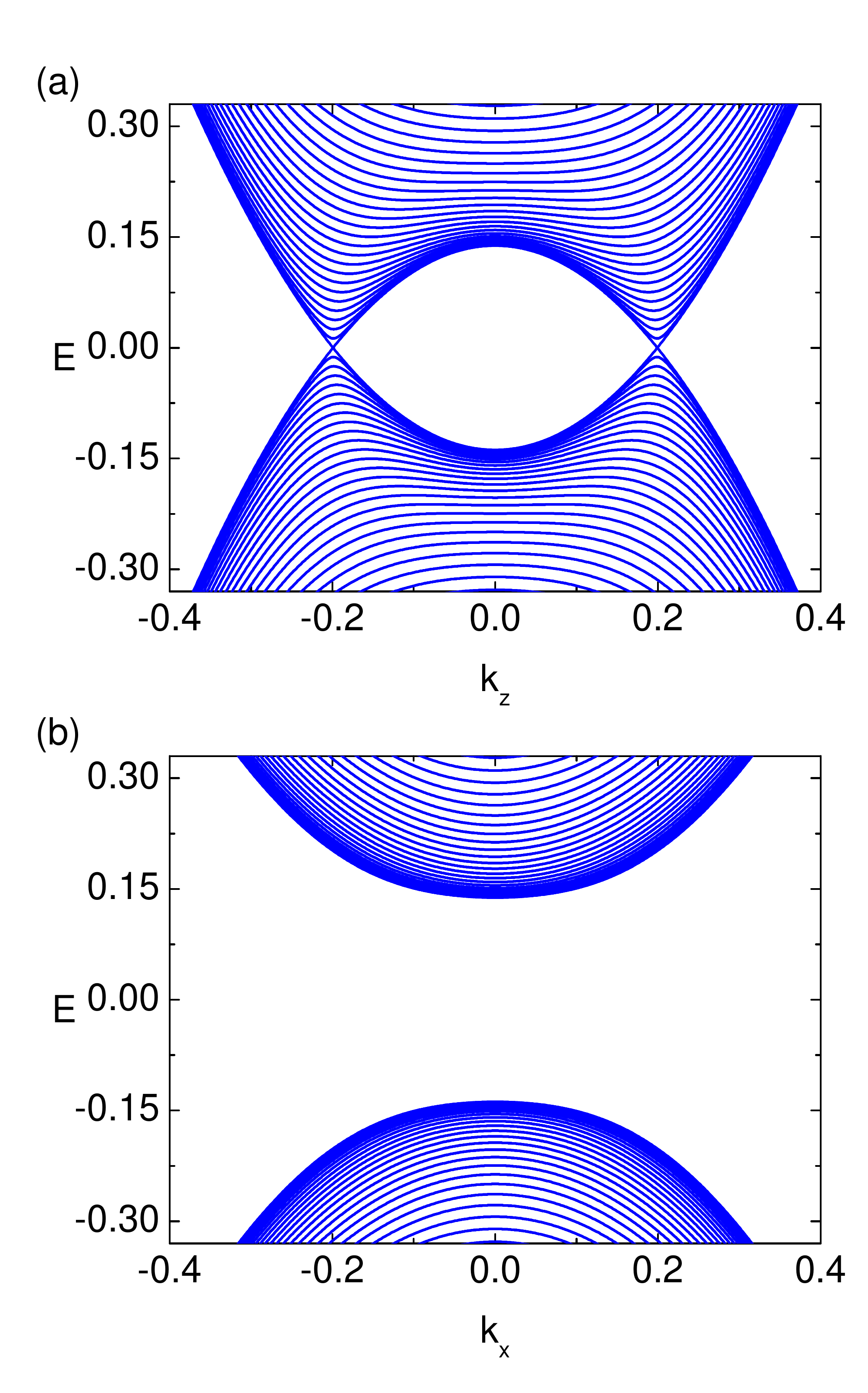}
\caption{Band structure of a Weyl superconductor plotted (a) along $k_z$-axis with $k_x=0$ and (b)  along $k_x$-axis with $k_z=0$. Periodic boundary conditions are applied in $x, z$-directions while the system is chosen to have $\bar{l}_y=500$ layers in $y$-direction. The parameters are listed below Eq.\ (\ref{spectrum}). } \label{fig2}
\end{figure}

As wtitten Hamiltonian Eq.\  (\ref{H}) is $\bm k \cdot \bm p$ in $x$-$y$ plane and tight-binding in $z$-direction. It will be useful to  apply lattice regularization. We use a simple cubic lattice  with lattice constant $a$ and replace $k_{x,y}\rightarrow \frac{1}{a} \sin ak_{x,y}$ and $k_{x,y}^2 \rightarrow \frac{2}{a^2}(1-\cos ak_{x,y})$. After partial Fourier transform in the $z$-direction, the Hamiltonian can be written as
\begin{equation}
H=\frac{1}{2} \sum\limits_{\bm k} \Psi_{\bm k}^\dagger \mathcal H_{\bm k} \Psi_{\bm k}, 
\end{equation}
with $\Psi_{\bm k} = (c_{\bm k, 1 \uparrow}, c_{\bm k, 1 \downarrow}, c_{\bm k, 2 \uparrow}, c_{\bm k, 2 \downarrow}, c^\dagger_{-\bm k, 1 \uparrow}, c^\dagger_{-\bm k, 1 \downarrow}, c^\dagger_{-\bm k, 2 \uparrow}, c^\dagger_{-\bm k, 2 \downarrow} )^T$ and 
\begin{multline} \label{Hk}
\mathcal H_{\bm k} = (m-4m'+2m'\cos k_xa+2m' \cos k_ya) s^z \tau^z + \\ 
t_d\sin k_za \sigma^y \tau^z + (t_s+t_d \cos k_za)\sigma^x\tau^z + \frac{\hbar v_F}{a} \sin k_ya s^x \sigma^z 
\\ - \frac{\hbar v_F}{a} \sin k_xa s^y \sigma^z \tau^z - \text{Im}\Delta s^y \tau^x - \text{Re}\Delta s^y\tau^y 
\end{multline}
The spectrum of $\mathcal H_{\bm k}$  reads
\begin{multline} \label{spectrum}
\epsilon_{\bm k, \pm}^2 = \frac{\hbar^2 v_F^2}{a^2}(\sin^2 k_xa + \sin^2 k_ya) +\\ \Big(m - 4m' + 2m' \cos k_xa + 2m' \cos k_ya \\ \pm \sqrt{t_s^2 + t_d^2 + 2t_st_d \cos k_za+|\Delta|^2} \Big)^2.
\end{multline}
We plot the spectrum in Fig.~\ref{fig2} for a system with $\bar{l}_y=500$ layers and open boundary conditions along the $y$-direction and  periodic boundary conditions in the other two dimensions. We set $\Delta=1$ and measure all other parameters in terms of $\Delta$. We take $m=10.26$, $m'=2.53$, $\hbar v_F/a =1$, $t_d=-4.79$, $t_s=14.86$, and the lattice constant is set to be $a=6\mathring{\text{A}}$. These values will also be used in our numerical simulations unless other values are specified.

\begin{figure*}[ht]
\includegraphics[width=16cm]{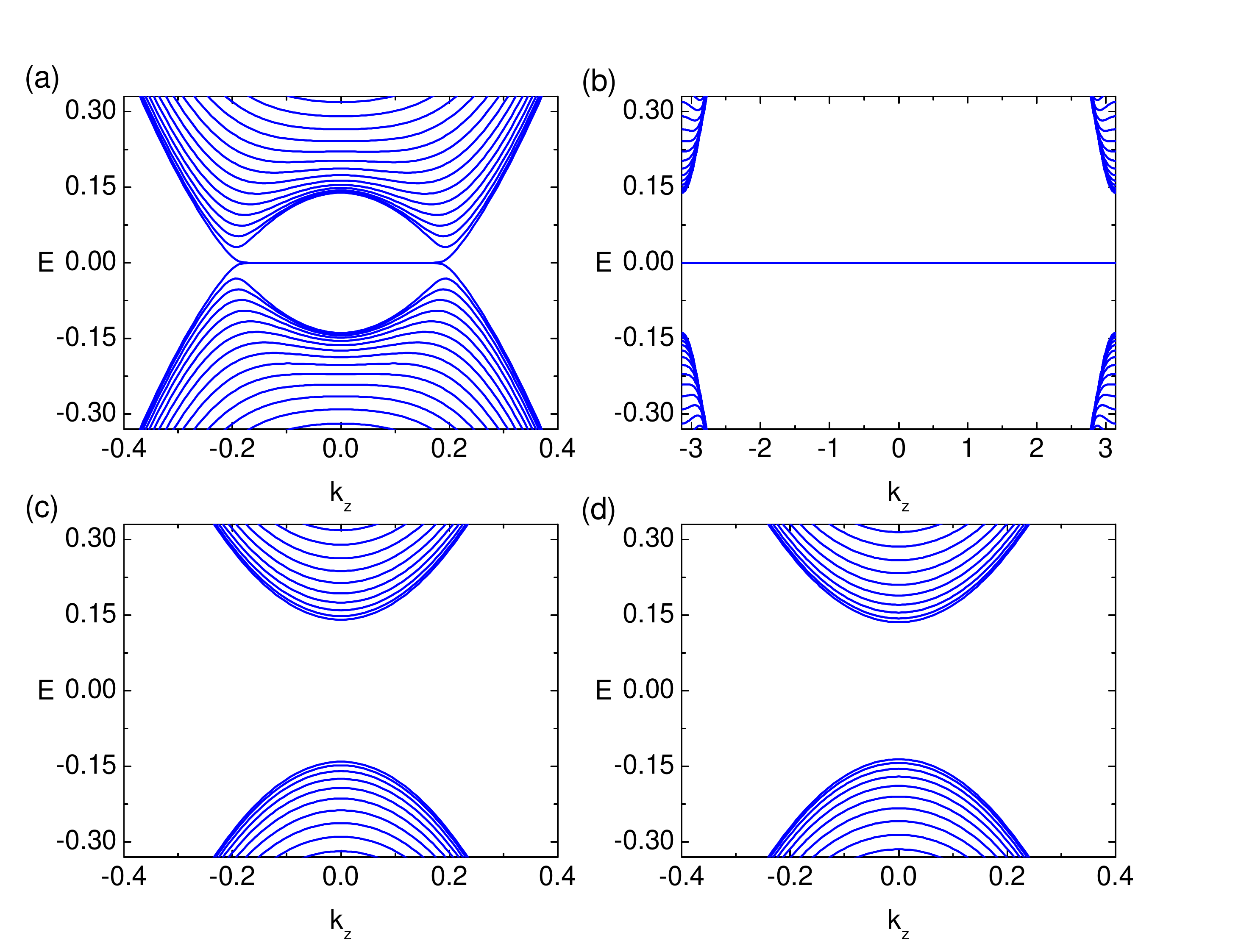}
\centering
\caption{ Band structure of a Weyl superconductor with open boundary conditions and $\bar{l}_y=150$ layers along the $y$-direction. All panels are plotted along $k_z$-axis  with $k_x=0$ and with patameters as in Fig.\ 2 .  (a) Weyl superconductor phase for $(m, \Delta) = (10.26,1)$. A Fermi arc connecting two Weyl points appears due to the chiral Majorana edge states of the effective $p_x+ip_y$  superconductors that emerge for fixed $k_z$ between the Weyl nodes. (b) Topological superconductor phase for $(m, \Delta) = (19.82,1)$. The increase of $m$ will separate two Weyl points and extend Fermi arc. When two Weyl points meet at Brillouin zone boundary, they annihilate and open up a SC gap but leave behind the Fermi arc extended over the whole BZ. (c) Trivial superconductor phase for $(m, \Delta) = (9.98,1)$. The decrease of $m$ makes two Weyl points meet at Brillouin zone center and annihilation and leads to the disappearance of the Fermi arc. (d) Trivial superconductor phase with $(m, \Delta) = (10.26,2.56)$. The increase of $\Delta$ is equivalent to decrease of $m$ and Weyl points again annihilate at the BZ center. }\label{fig3}
\end{figure*}

Without loss of generality, we have assumed $m, m'>0$ in the following discussion. Thus, the sector $\epsilon_{\bm k, +}$ is fully gapped while $\epsilon_{\bm k,-}$ can be gapless when
\begin{equation} \label{m}
\sqrt{(|t_s| - |t_d|)^2 +|\Delta|^2}<m<\sqrt{(|t_s| + |t_d|)^2 +|\Delta|^2}
\end{equation}
If Eq.\  (\ref{m}) holds, nondegenerate quasiparticle bands exhibit a point node  at $\bm k_W=(0, 0, \eta Q)$ with
\begin{equation}\label{Qa}
Qa=\cos^{-1} \bigg(\frac{m^2 - t_s^2 - t_d^2 - |\Delta|^2}{2t_st_d}\bigg)
\end{equation}
and $\eta = \pm 1$. As expected, the system is a Weyl superconductor.

To understand the low-energy physics better, we introduce $2\times 2$ auxiliary matrices
\begin{multline} 
D_{\bm k, \pm} = \frac{\hbar v_F}{a}\sin k_ya\kappa^x - \frac{\hbar v_F}{a}\sin k_xa\kappa^y +\\ (M_{k_z, \pm} - 4m' + 2m' \cos k_xa +2m' \cos k_ya) \kappa^z
\end{multline}
where $ \bm \kappa$ are Pauli matrices in transformed Nambu space and
$$
M_{k_z, \pm} = m  \pm \sqrt{t_s^2 + t_d^2 + 2t_st_d \cos k_za+|\Delta|^2}
$$
As $\epsilon_{\bm k, \pm}$ is also the dispersion for $D_{\bm k, \pm}$, the low-energy physics of Eq.\ (\ref{Hk}) may be understood by studying $D_{\bm k, -}$ because there always exists a unitary transformation $U$ that can block diagonalize $\mathcal H_{\bm k}$ 
$$
U^{-1} \mathcal H_{\bm k} U = \text{diag} (D_{\bm k, -}, D_{\bm k, -}, D_{\bm k, +}, D_{\bm k, +})
$$
For fixed $k_z$ value, we rewrite $D_{\bm k,-}$ in a $\bm k \cdot \bm p$ fashion,
\begin{equation} \label{D-}
D_{\bm k \cdot \bm p}^- = \begin{pmatrix}
-m' a^2 (k_x^2+k_y^2) + \mu_{\text{eff}} && i\hbar v_F (k_x - ik_y)\\-i\hbar v_F (k_x + ik_y) && m' a^2 (k_x^2+k_y^2) - \mu_{\text{eff}}
\end{pmatrix}
\end{equation}
We notice that $D_{\bm k \cdot \bm p}^-$ can be regarded as describing  a $p_x + i p_y$ superconductor with an effective chemical potential $\mu_{\text{eff}}=M_{k_z, -}$. It is characterized by Chern number
\begin{equation} \label{chern}
C=\frac{1}{2} [\text{sgn} (\mu_{\text{eff}}) + \text{sgn} (m')].
\end{equation}
If Eq.\  (\ref{m}) holds, for those $k_z$'s that satisfy $\mu_{\text{eff}}=M_{k_z, -}>0$, this SC is in the weak pairing phase with Chern number $C=1$. As a result, for each of such $k_z$'s, there exist counter propagating chiral Majorana states on boundaries  open in the  $y$-direction. Therefore, the edge states of Eq.\ (\ref{H}) are Majorana-Fermi arcs as illustrated in Fig.~\ref{fig3}(a).

To understand the phases of this model consider  a value of $m$ that satisfies Eq.\  (\ref{m}) with fixed $\Delta$.  Now  increase it such that $m>\sqrt{(|t_s|+|t_d|)^2+|\Delta|^2}$; according to Eq.\ (\ref{spectrum}), our Weyl superconductor will be gapped into a topologically superconducting phase, whose spectrum is shown in Fig.~\ref{fig3}(b). It exhibits a surface mode because the Chern number is still $C=1$ for all $k_z$. On the other hand, if $m$ is decreased to $m<\sqrt{(|t_s|-|t_d|)^2+|\Delta|^2}$, the system enters a trivial superconducting phase with no edge modes, as shown in Fig.~\ref{fig3}(c). If $m$ is fixed to a value obeying Eq.\ (\ref{m}), but $|\Delta|$ is gradually increased, eventually, $m$ is overwhelmed by $\sqrt{(|t_s|-|t_d|)^2+|\Delta|^2}$ and the system becomes a trivial superconductor, as indicated by Fig.~\ref{fig3}(d).

\begin{figure}[ht]
\includegraphics[width=6cm]{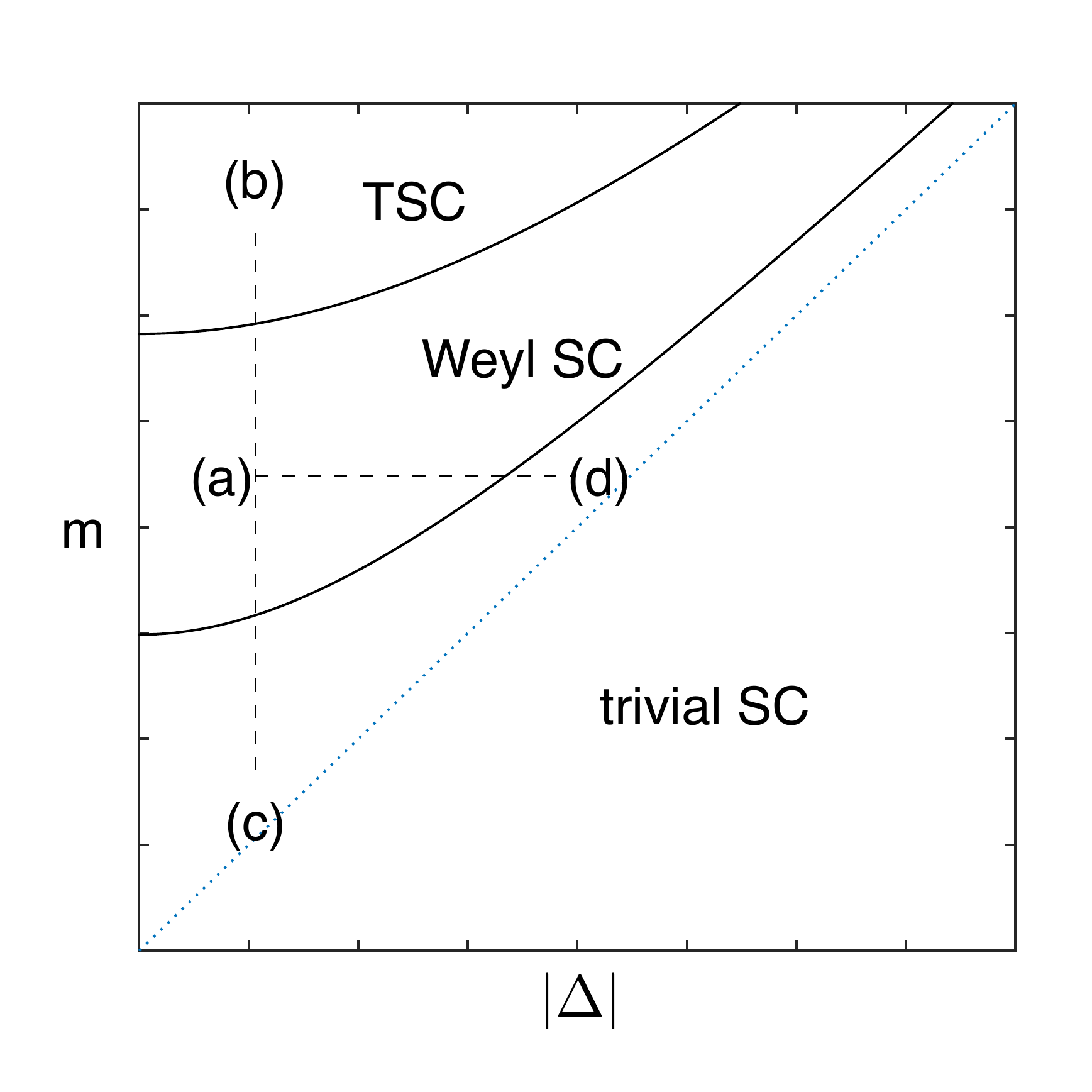}
\caption{Phase diagram of the Weyl superconductor described by Hamiltonian (2) in terms of $(m, |\Delta|)$ with labels (a)-(d) correspond to spectra shown in Fig.~\ref{fig3}(a)-(d). The two black curves mark the phase boundaries given in Eq.\ (\ref{upper_bound}) and Eq.\ (\ref{lower_bound}). The dotted line indicates the asymptote for the two phase boundaries.}\label{fig4}
\end{figure}

Based on the above considerations, we plot the global phase diagram of our Weyl superconductor in $|\Delta|-m$ plane in Fig.~\ref{fig4}. The phase boundaries are given by two hyperbolas,
\begin{gather}
m^2-|\Delta|^2 = (|t_s|+|t_d|)^2 \label{upper_bound},
\\
m^2-|\Delta|^2 = (|t_s|-|t_d|)^2 \label{lower_bound}.
\end{gather}
Above the upper bound Eq.\  (\ref{upper_bound}), the multilayer is a topological superconductor, which can be viewed as a stack of 2D $p_x+ip_y$ superconductors. These are known to posess counter propagating chiral Majorana edge modes on a pair of parallel boundaries. Since switching off the superconductivity will give a 3D quantum anomalous Hall (QAH) insulator \cite{burkov2011}, the multilayer topological superconductor structure may be refered to as ``3D QAH superconductor''.  Below the lower bound Eq.\  (\ref{lower_bound}), the multilayer is a trivial superconductor while  between them it is a Weyl superconductor.

\begin{figure*}
\includegraphics[width=16cm]{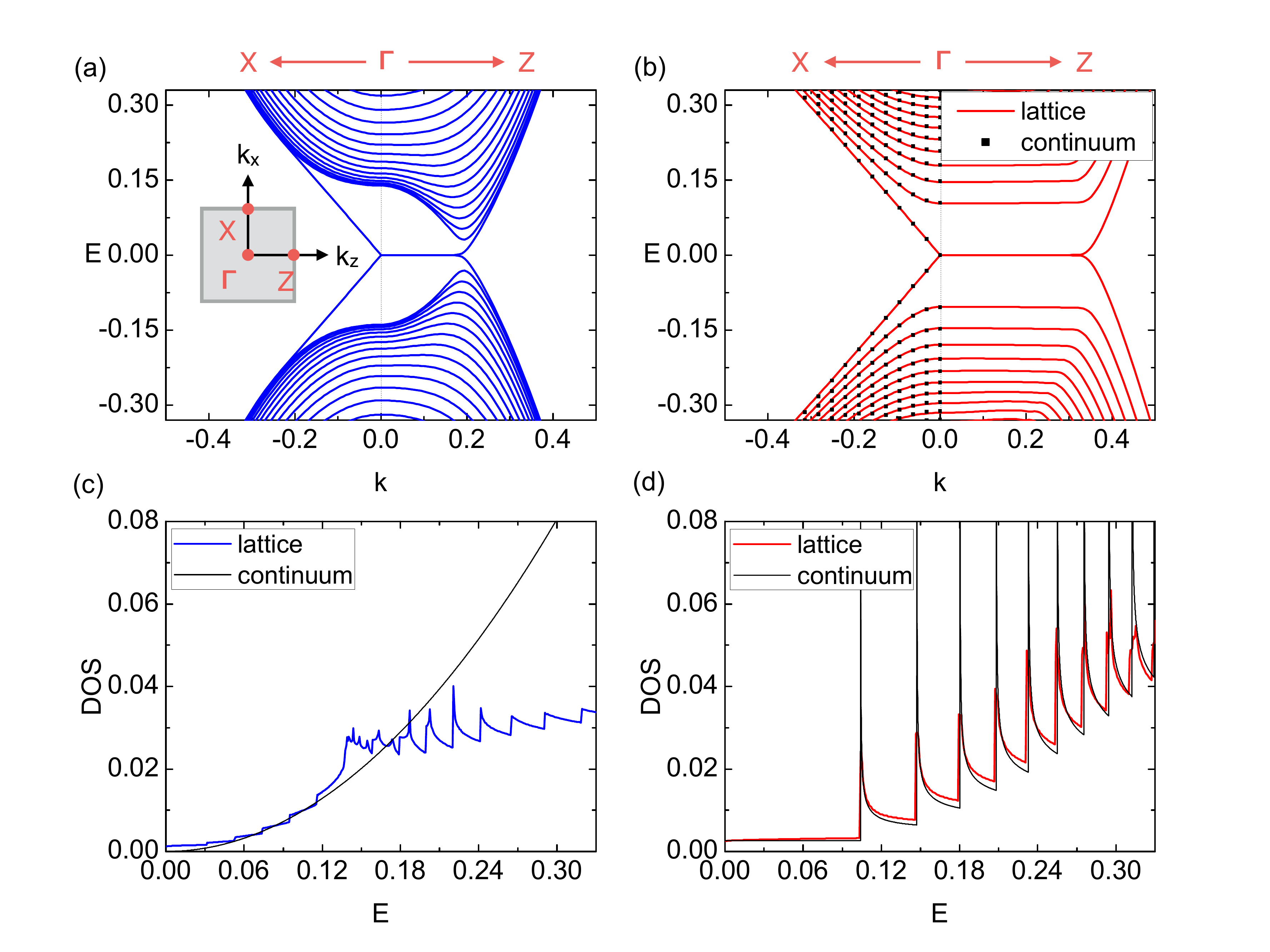}
\caption{Energy spectra and DOS for our  Weyl superconductor with open boundaries and $\bar{l}_y=150$ along the  $y$-direction and periodic along $x$ and $z$. (a) The spectrum of undeformed system; the flat band at zero energy is the Fermi arc. (b) The spectrum of a bent Weyl superconductor as shown in Fig.~\ref{fig1}(b) with $\varepsilon=8\%$ corresponding to a pseudo-magnetic field $b=10.45 \text{T}$. For both (a) and (b) the spectrum is plotted along $X$-$\Gamma$-$Z$ as shown in the inset. For comparison, energy levels Eq.\ (\ref{En}) are overlain as black dots. (c) DOS of the unstrained sample (blue curve)  compared to the ideal $\sim E^2$ DOS expected for a  massless Dirac fermion in continuum  (black parabola). (d) DOS of the strained system (red curve) compared to  DOS calculated for ideal Dirac-Landau levels with $b=10.45 \text{T}$.
}\label{fig5}
\end{figure*}

\section{Strain-induced gauge field}
\label{Strain induced gauge field}
In the preceding section, we studied electronic structure and phase diagram of multilayer model of Weyl superconductor. In this section, we will understand how the electronic structure is changed under generic strain. 

When elastic strain distorts the lattice, the chemical bonds are stretched and compressed. Orbital orientations are also rotated, making the symmetry-prohibited hoppings now non-zero. For our purposes, the most important modification comes from the replacement of hopping amplitudes along $z$-direction \cite{pikulin2016, liu2017, shapourian2015}
\begin{equation} 
\begin{split}
 t_d  \sigma ^ \pm \rightarrow  t_d(1-w_{33}) \sigma ^ \pm - i \frac{\hbar v_F}{a}& w_{31} s^y \sigma^z + i \frac{\hbar v_F}{a} w_{32} s^x \sigma^z \\ 
t_s \rightarrow t_s(1 -&w_{33}) \label{t}
\end{split}
\end{equation}
with the strain tensor $w_{ij}=\frac{1}{2}(\partial_i u_j + \partial_j u_i)$, where $u_j$ is the $j$-th component of the displacement vector $\bm u$. Under such hopping parameter substitution, the Hamiltonian in Eq.\ (\ref{Hk}) is changed to 
\begin{equation}\label{strain_Hk}
\tilde{\mathcal{H}}_{\bm k} = \mathcal{H}_{\bm k} + \delta \mathcal{H}_{\bm k}
\end{equation}
 where the correction due to strain is
\begin{multline} 
\delta \mathcal{H}_{\bm k} = -(t_s w_{33} + t_d w_{33} \cos k_za ) \sigma^x \tau^z -  t_d w_{33} \sigma^y \tau^z \sin k_za 
\\- \frac{\hbar v_F}{a} w_{31} s^y \sigma^z \tau^z \sin k_za + \frac{\hbar v_F}{a} w_{32} s^x \sigma^z \sin k_za.
\end{multline}
To understand the effect of strain on the low-energy physics we consider $\tilde{\mathcal{H}}_{\bm k}$ in the vicinity of Weyl points $\bm k_W = (0, 0, \eta Q)$ and expand
\begin{equation} \label{strain_H}
\tilde{\mathcal{H}}_{\bm k_W + \bm q} = \mathcal{H}_{\bm k_W} + h_{\bm q}+ \delta \mathcal{H}_{\bm k_W} + O(q^2) + O(q)Q(w_{ij}),
\end{equation}
 where we only keep terms up to the linear order in  momentum $\bm q$ and the strain  tensor $w_{ij}$. We find
 \begin{multline} \label{linearized Hk}
 h_{\bm q}= \bm q \cdot \nabla_{\bm k} \mathcal{H}_{\bm k} |_{\bm k =\bm k_W} = 
-  \hbar v_Fq_x s^y \sigma^z \tau^z + \hbar v_F  q_y s^x \sigma^z 
\\
- \eta t_d q_za\sin Qa  \sigma^x\tau^z + t_d q_za\cos Qa  \sigma^y \tau^z.
\end{multline}
If we assume constant strain $\nabla_{\bm r }w_{ij}=0$, the spectrum of Eq.\ (\ref{strain_H}) can be found
\begin{multline} \label{strain_spectrum}
\tilde{\epsilon}_{\bm k, \pm}^2 = \frac{\hbar^2 v_F^2}{a^2} \big[(q_x + \eta w_{31} \sin aQ)^2 +
(q_y + \eta w_{32} \sin aQ)^2\big]
\\
+\bigg[m \pm m \mp \frac{\eta t_s t_d \sin aQ}{m} \bigg(q_za + \eta w_{33} \frac{m^2-|\Delta|^2}{t_s t_d \sin aQ}\bigg)\bigg]^2
\end{multline}
From this we can extract  the magnitudes of the Fermi velocity components
\begin{equation} \label{vF}
(v_x,v_y,v_z)=\Big(v_F, v_F, \Big| \frac{t_s t_d \sin aQ}{m}\Big| \Big).
\end{equation}
In addition we observe that the  effect of constant distortion is to translate the Weyl points in the momentum space by  
\begin{equation} \label{A}
e \mathcal{A} = - \frac{ \eta \hbar}{a} \Big(w_{31} \sin Qa , w_{32} \sin Qa , w_{33} \frac{m^2-|\Delta|^2}{t_st_d \sin Qa}\Big)
\end{equation}
Clearly, the vector $\mathcal{A}$ can be understood as the gauge potential of a strain-induced chiral magnetic field. 

To gain additional insight into this result, one can project $h_{\bm q}+ \delta \mathcal{H}_{\bm k_W}$ onto the Hilbert space spanned by 4 zero-energy eigenvectors of $\mathcal{H}_{\bm k_W}$. The resulting Hamiltonian has massless  Dirac form with the strain entering via minimal substitution. The strain induced chiral gauge potential can then be extracted and is consistent with Eq.\  (\ref{A}). Details about this approach are given  in Appendix~\ref{Low-energy Weyl Hamiltonian}. 

In most cases we expect  $Qa \ll 1$. In this limit  the $z$-component of $\mathcal{A}$  given in Eq.\  (\ref{A}) scales as $1/aQ$ but $x(y)$-components scale as $aQ$. Thus only $\mathcal{A}_z$ will be considered in the following. 

We obtained $\mathcal{A}$ above by assuming constant distortion $w_{ij}$. However, we are interested in the bending deformation as shown in Fig.~\ref{fig1}(b) where $w_{ij}$ is obviously space-dependent. We argue that even in this case the strain effect can be treated as a gauge field expressed in Eq.\  (\ref{A}), as long as it varies slowly on the lattice scale. To support our argument, we first find the expression for $w_{33}$ when the system is bent and then numerically calculate the spectra of the tight-binding Hamiltonian with this distortion.

The bending deformation can be characterized by an small angle $\theta = a/\rho$ where $\rho$ is the radius of the circular bend. The lattice constant of the outermost $y$-direction layer is then $a+\delta a$ with $\delta a = \frac{1}{2} \bar{l}_y a \theta$. Here $\bar{l}_y$ is the number of layers in $y$-direction. Thus
\begin{equation}
\rho = \frac{\bar{l}_y a}{2\delta a/a} = \frac{\bar{l}_y a}{\varepsilon}
\end{equation}  
with $\varepsilon = 2 \delta a/a$ the bending parameter used in the numerics. Now if we consider a generic $y$-direction layer, its lattice constant will change by $\delta a(y) = (y-\bar{l}_y a/2)\theta$. Then for a point with coordinate $z$ on this layer, its $z$-direction displacement is $ u_3 = \frac{z}{a} (y-\bar{l}_y a/2)\frac{2\delta a}{\bar{l}_y a}$. Thus, 
\begin{equation}
w_{33} = \frac{\partial u_3}{\partial z} = (y-\bar{l}_y a/2)\frac{2\delta a}{a^2 \bar{l}_y} =(y-\bar{l}_y a/2)\frac{\varepsilon}{\bar{l}_y a}.
\end{equation}
Therefore, we expect a pseudo-magnetic field
\begin{equation} \label{pb}
\bm b = \partial_y \mathcal{A}_z \hat x= - \eta \frac{\hbar}{ea^2}  \frac{m^2-|\Delta|^2}{t_s t_d \sin Qa} \frac{\varepsilon}{\bar{l}_y} \hat x
\end{equation}
 Such pseudo-magnetic field will give rise to Dirac-Landau levels at energies
\begin{equation} \label{dl}
\tilde{\epsilon}_n(\bm k) =\pm \sqrt{\hbar^2v_x^2k_x^2 + 2n \Big| \frac{eb}{\hbar} \hbar v_y \hbar v_z \Big|}
\end{equation}
for all integers $n \neq 0$ and $\tilde{\epsilon}_0(k_x)= -\hbar v_F k_x$ as the zeroth Landau levels for both valleys. In view of Eq.\  (\ref{pb})  we get
\begin{equation}\label{En}
\tilde{\epsilon}_n(\bm k) = \pm \sqrt{\hbar^2v_x^2k_x^2 + 2 n \frac{\varepsilon}{\bar{l}_y} \frac{m^2 - |\Delta|^2}{m} \frac{\hbar v_y}{a}}.
\end{equation}

We have numerically checked Eq.\ (\ref{En}) by applying hopping substitutions Eq.\ (\ref{t}) in the multilayer Hamiltonian Eq.\ (\ref{H}) with $\bar{l}_y=150$,  as summarized in Fig.~\ref{fig5}. Indeed we observe that  the  Dirac-Landau levels in Eq.\ (\ref{dl}) capture the features of the low-energy spectrum of the Weyl superconductor multilayer. For comparison we also plot the spectrum and DOS for the unstrained system and show the result in Fig.~\ref{fig5}(a,b).

For the sake of completeness we in addition calculate the spectrum of our model Weyl superconductor  in the presence of the magnetic field ${\bf B}\parallel \hat{z}$ and the Abrikosov vortex lattice. This is summarized in Appendix \ref{Vortex lattice}. We find that all bands become completely flat Landau levels in the $x$-$y$ plane. The zeroth Landau level, which is associated with Weyl nodes before $\bf B$ is switched on, is still linearly dispersive along the $z$-direction in the vicinity of nodes. In contrast, it is well known that the  magnetic field does not lead to flat Landau levels in $d_{x^2-y^2}$ superconductors. This is because the spatially varying supercurrent in the vortex lattice strongly scatters the Bogoliubov quasiparticles  \cite{franz2000}. The difference between the $d_{x^2-y^2}$ and 3D Weyl superconductors  has been recently elucidated in Ref. \cite{pacholski1} with which our results are in accord. In short, the zeroth Landau level cannot be scattered by vortices due to the protection of Weyl node chirality, which is a topologically nontrivial and unique feature of the  Weyl superconductor.

\section{Longitudinal thermal conductivity}
\label{Longitudinal thermal conductivity}
In Section \ref{Strain induced gauge field}, we found that a bending deformation results in Dirac-Landau levels for Bogoliubov  quasiparticles, which unlike those in Weyl semimetals, are charge neutral on average.  Therefore, Shubnikov-de Haas quantum oscillation discussed in Ref. \cite{liu2017} cannot be observed in Weyl superconductors. But Bogoliubov quasiparticles do carry heat, making thermal transport measurements, such as the thermal Hall effect, possible \cite{murray2015, sumiyoshi2013, wang2011}.

We will evaluate in this Section the longitudinal thermal conductivity $\kappa_{xx}$ as a function of pseudo-magnetic field  and show that $\kappa_{xx}$ exhibits oscillations periodic in $1/b$. In our analysis above the chemical potential $\mu$ of TI layers was assumed to lie at the Weyl point. Here we will tune $\mu$ away from this ``neutrality point'' to obtain a finite size Fermi surface in order to observe quantum oscillation in $\kappa_{xx}$. For $\mu\neq 0$  Eq.\ (\ref{Hk}) will be modified by an extra term $\delta \mathcal{H}_{\bm k} = -\mu \tau^z$. This term lifts the 2-fold degeneracy in the spectrum Eq.\ (\ref{spectrum}). Perturbative analysis  detailed in Appendix~\ref{Weyl superconductor with chemical potential} shows that the resulting spectrum can be well approximated by split Dirac-Landau levels of the form
\begin{equation} \label{biased Dirac-Landau}
E_n(\bm k) =\pm \sqrt{\hbar^2 v_x^2 k_x^2 + 2n \Big| \frac{eb}{\hbar} \hbar v_y \hbar v_z\Big|} \pm \mu,
\end{equation}
\begin{figure}[t]
\includegraphics[width=7cm]{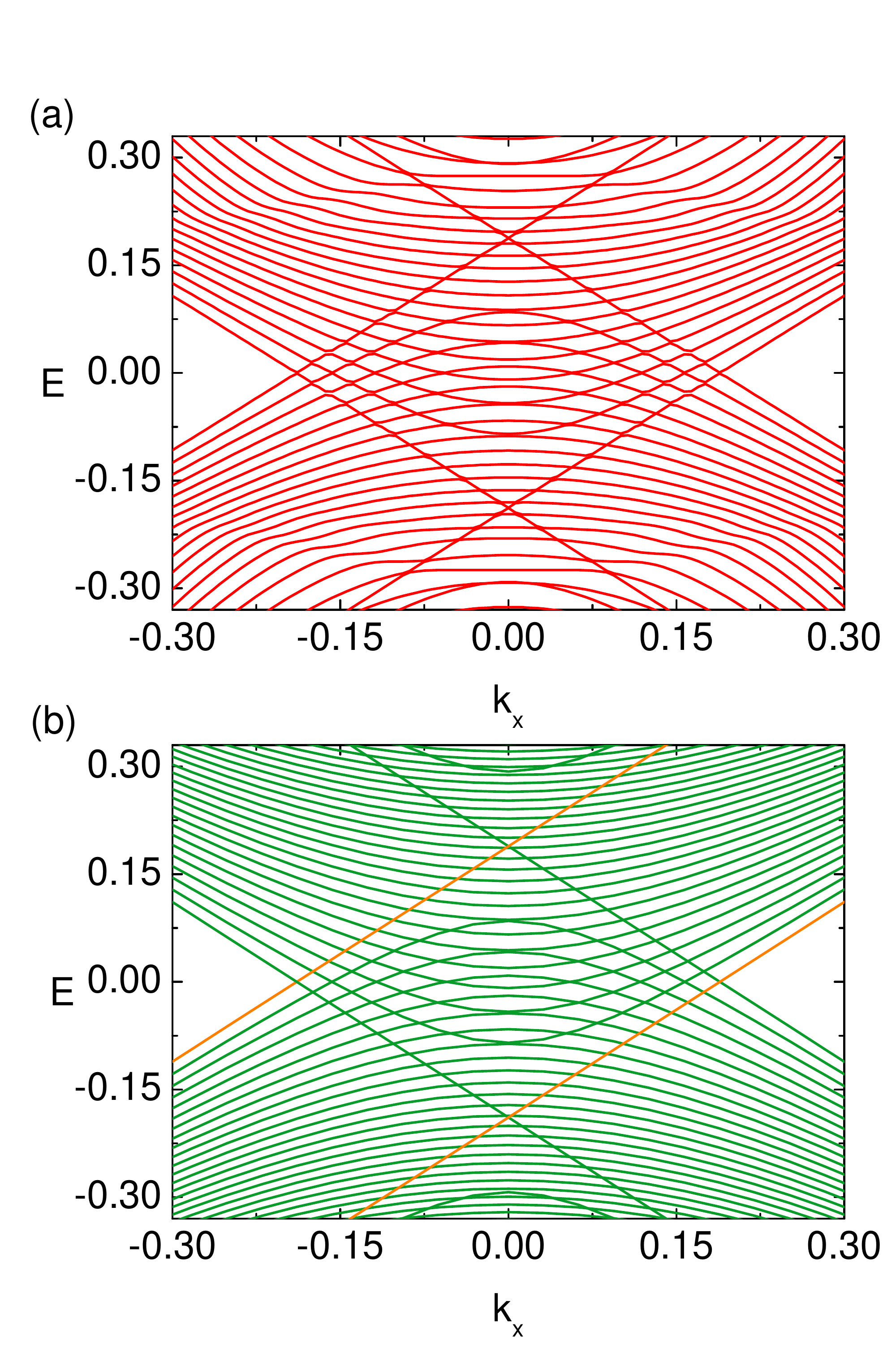}
\caption{Energy spectrum of the Weyl SC with the chemical potential of the TI layers tuned away from the surface Dirac points to $\mu = 0.19$. (a) Quasiparticle spectrum calculated from the lattice model Eq.\ (\ref{strain_Hk}). It is worth noting that only the left moving chiral mode is due to the Landau quantization while the other is a surface mode. (b) Quasiparticle spectrum predicted by Eq.\ (\ref{biased Dirac-Landau}). To compare with the first panel, the chiral modes (orange lines) due to the surface states have been added manually.}
\label{fig6}
\end{figure}
illustrated in Fig.~\ref{fig6}(b) . The corresponding DOS at the chemical potential is
\begin{multline} \label{DOS}
D(0) 
=\frac{L_yL_z}{2\pi l_B^2} \sum\limits_n L_x \sum\limits_\pm \int \frac{dk_x}{2 \pi}\delta(E_n(\bm k)),
\\
= \frac{V}{2\pi^2l_B^2} \frac{2}{\hbar v_x} \sum\limits_n  \sqrt{\frac{\mu^2}{\mu^2-2n |\frac{eb}{\hbar} \hbar v_y \hbar v_z |}},
\end{multline}
where $l_B=\sqrt{\hbar/eb}$ is the magnetic length. The thermal conductivity can be computed using the Boltzmann equation approach \cite{ong1999, bardeen1959, tewordt1962} and reads
\begin{multline}\label{kappaxx1}
\bm \kappa=  \frac{1}{T}\sum\limits_n\sum\limits_{\bm k}E_n^2(\bm k) 
\tau_n(\bm k) \bm v_n(\bm k) \bm v_n(\bm k) \bigg(- \frac{\partial f}{\partial E_n} \bigg),
\end{multline}
where $E_n(\bm k)$ is the quasiparticle energy, $\bm v_n(\bm k) = \frac{1}{\hbar} \nabla_{\bm k} E_n(\bm k)$ is the associated velocity, $\tau_n(\bm k)$ is the corresponding scattering time, and $f(E_n)=(e^{E_n/k_BT}+1)^{-1}$. For our purposes, it is useful  to rewrite the thermal conductivity (\ref{kappaxx1}) in the low-$T$ limit through the Sommerfeld expansion as explained in Appendix~\ref{Thermal conductivity}. To leading order one obtains
\begin{equation} \label{Wiedemann-Franz}
\bm \kappa =  \frac{\pi^2k_B^2T}{3}\sum\limits_n\sum\limits_{\bm k} \tau_n(\bm k) \delta(E_n(\bm k)) \bm v_n(\bm k) \bm v_n(\bm k).
\end{equation}
The scattering rate can be approximated by Fermi's golden rule (see Appendix~\ref{Thermal conductivity}) as
\begin{equation} \label{tau}
\tau_n^{-1}(\bm k)\big|_{E_n(\bm k)=0} = \frac{2\pi}{\hbar} n_{\text{imp}} C_{\text{imp}} D(0)
\end{equation}
where $n_{\text{imp}}$ and $C_{\text{imp}}$ denote the impurity concentration and the scattering potential strength, respectively. The longitudinal thermal conductivity then becomes
\begin{equation} \label{kappa}
\kappa_{xx}(b) = \kappa_{xx}(0) \frac{\sum\limits_n \sqrt{\frac{ \mu^2 -2n |\frac{eb}{\hbar} \hbar v_y \hbar v_z |}{ \mu^2}}}{\sum\limits_n  \sqrt{\frac{\mu^2}{\mu^2-2n |\frac{eb}{\hbar} \hbar v_y \hbar v_z |}}}
\end{equation}
with the zero-field thermal conductivity 
\begin{equation}
\kappa_{xx}( 0)=\frac{\pi^2k_B^2T}{3} 
\frac{v_x^2}{\frac{2\pi}{\hbar} n_{\text{imp}}C_{\text{imp}}}.
\end{equation}
Fig.~\ref{fig7} shows our results for DOS and $\kappa_{xx}(b)$ calculated from the approximate analytical formulas Eqs.\ (\ref{DOS}) and (\ref{kappa}), and based on the full lattice calculation using  Eqs.\ (\ref{DOS}) and (\ref{Wiedemann-Franz}). They agree well and exhibit pronounced quantum oscillations periodic in $1/b$.  

We note that due to Landau quantization, thermal conductivity quantum oscillations in the $x$-direction are most pronounced, while in the other directions, quantum oscillations are expected  to be weaker. Based on our results for the electronic structure in Fig.~\ref{fig5}(b), the $z$-direction drift velocity of Bogoliubov quasiparticles is nonzero only at the edges of bands ($k_za \sim 0.3$). In contrast, the $x$-direction drift velocity is nonzero for almost all momenta. Therefore, $\kappa_{zz}$ should be small and its quantum oscillations are weaker than those in $\kappa_{xx}$. In y-direction, quasi-particle wave functions are Gaussian-localized with the characteristic decay length $\sqrt{\hbar v_y/ebv_z} \sim l_B$ and localization centers $2\pi \nu l_B^2/L_z$ with $\nu=1,2,\cdots,L_z/a$. The localization makes transport difficult unless the localization center is pumped across the system when $b$ varies. Therefore, we do not expect pronounced quantum oscillations along the $y$-direction.

It is also worth noting that quasiparticle thermal conductivity can be obscured in real materials by phonons because phonons also carry heat. For temperature $T \ll T_c$, the thermal conductivity of acoustic phonons follows Debye $T^3$ law $\kappa_{ph}^A\sim T^3$. The less dominant optical phonon thermal conductivity is $\kappa_{ph}^O\sim \frac{1}{T^2} \exp(-\frac{1}{T})$. At low temperatures both will be overwhelmed by quasiparticle contribution. At higher temperatures, the phonon contribution $\kappa_{ph}=\kappa_{ph}^A+\kappa_{ph}^O$ can dominate over the quasiparticle thermal conductivity but quantum oscillations should remain visible. We do not expect $\kappa_{ph}$ to show quantum oscillations because phonons are bosonic excitations.
\begin{figure}[t]
\includegraphics[width=9cm]{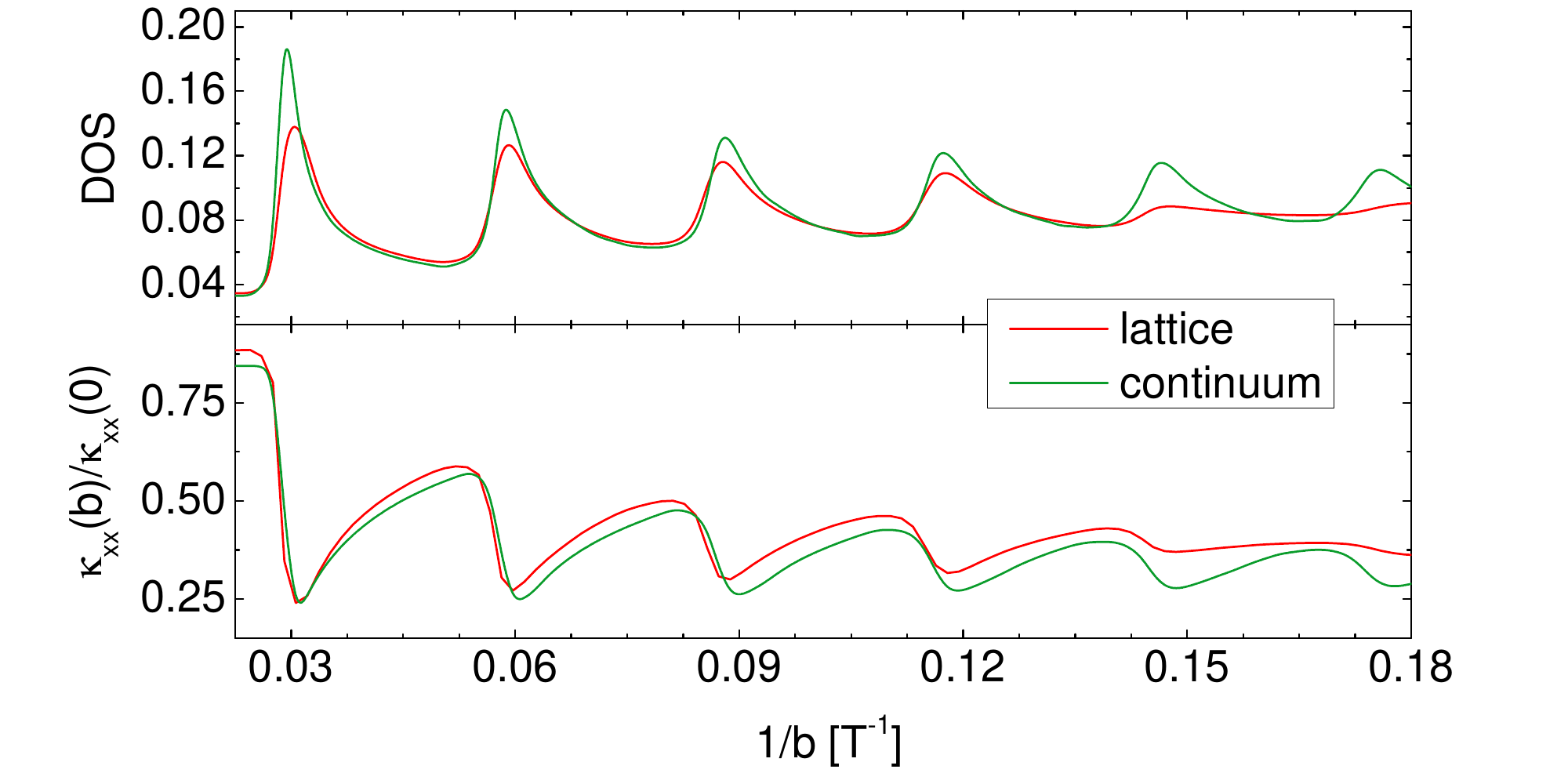}
\caption{Strain-induced quantum oscillation in a Weyl semimatal. The upper panel shows oscillations in DOS as a function of inverse strain strength expressed as $1/b$ at zero-energy. The lower panel shows oscillations in the longitudinal quasiparticle thermal conductivity $\kappa_{xx}$. To simulate the effect of disorder, all data are broadened by convolving in energy with a Lorentzian with width $\delta_{\epsilon} = 1.67\times 10^{-3}$. }
\label{fig7}
\end{figure}

\section{Conclusions}
\label{Conclusion}
In this work, we studied a minimal model for a Weyl superconductor with a single pair of Weyl points based on the Meng-Balents layer construction. A Majorana-Fermi arc appears and connects the two Weyl points if a pair of boundaries are open. This arc can be understood as being formed of two counter propagating chiral Majorana modes at the edge of an effective topological $p_x+ip_y$ SC that results from fixing one component of the momentum in the 3D Hamiltonian describing the original Weyl SC. The phase diagram shows that the Weyl SC phase appears intermediate between a fully gapped trivial superconducting phase and a topological superconductor phase. These features  elucidate similarities between Weyl superconductors and Weyl semimetals.

In the low-energy sector of the theory, we showed that elastic strain acts as a chiral gauge potential incorporated in the Weyl Hamiltonian through standard minimal  substitution. Therefore, similar to graphene and Weyl semimetals, strain can mimic the effect of real physical magnetic field in the Weyl superconductor. One important difference is that the strain-induced pseudo-magnetic field is not subject to the Meisner effect. Remarkably, this fact allows the pseudo-magnetic field to Landau quantize the spectrum of Bogoliubov quasiparticles instead of being expelled from the sample or creating Abrikosov vortex lattice as would be the case for physical magnetic field $B$.

Landau quantization generates pronounced quantum oscillations that can be  observed by quasiparticle spectroscopy and by longitudinal thermal conductivity. These quantum oscillations occur deep in the superconducting state and are thus fundamentaly different from various theoretical proposals and experimental results that pertain to mixed and normal states of superconductors.

To experimentally test our proposal we require a Weyl superconductor. Such can be in principle artificially engineered through the Meng-Balents construction or can occur naturally in a suitable crystalline solid. Currently, there are roughly twenty different nodal superconductors known to science \cite{schnyder2015}. One of the most promising candidates may be Cu$_x$Bi$_2$Se$_3$ \cite{fu2010, yang2014}. Nuclear magnetic resonance experiments \cite{matano2016} revealed broken spin rotational symmetry in Cu$_x$Bi$_2$Se$_3$, suggesting the superconducting gap structure to be either $\Delta_{4x}$ where the nodes appear due to the protection of mirror symmetry or $\Delta_{4y}$ where small gaps (or nodes) are expected. Recent experimental results on the specific heat \cite{yonezawa2016} of Cu$_x$Bi$_2$Se$_3$ are consistent with nematic superconductivity and favor $\Delta_{4y}$ pairing structure. Unfortunately, the gap minima and nodes cannot be straightforwardly differentiated based on the reported specific heat data alone. However, symmetry and energetic considerations \cite{fu2014, venderbos2016} suggest gap minima in nematic superconductivity. Another promising candidate is Nb$_x$Bi$_2$Se$_3$, whose low temperature penetration depth exhibits quadratic temperature dependence characteristic of linearly dispersing point nodes in three dimensions \cite{smylie2016}. This is consistent with Nb$_x$Bi$_2$Se$_3$ being a Weyl  superconductor. Although it is too early  to draw a firm conclusion regarding the pairing state of the candidate materials, the existing experimental data give hope that Cu$_x$Bi$_2$Se$_3$ and Nb$_x$Bi$_2$Se$_3$ could eventually be identified as 3D  Dirac or Weyl superconductors.

The second requirement is that the candidate material be sufficiently flexible to allow  a few percent elastic deformation in order to generate a sufficiently strong pseudo-magnetic field. The candidate material should be prepared in a nanoscale thin film geometry in order to maximize its flexibility. To the best of our knowledge, detailed data on the mechanical properties of Cu$_x$Bi$_2$Se$_3$ is lacking and further experimental work is needed to determine whether or not this could be a suitable material.

There are several future directions that might be interesting to pursue based on our current work. The first is to test other properties associated with pseudo-Landau levels. Recent work on the fractional Josephson effect in strained 2D graphene superconductor \cite{lee2017} motivates the interest in studying a similar effect in one dimension higher using strained Weyl superconductor. The second lies in the study of the chiral anomaly, chiral magnetic effect, and gravitational anomaly with strain-induced gauge field.

\begin{acknowledgments}
The authors are indebted to T. Meng, A. Tsuruta, Y. Yanase, T. Matsushita, Z. Shi, and A. Chen for illuminating discussions. TL thanks JREP program under Topo-Q network offered by the "Topological Materials Science" project for the support at the initial stage of this work. SF is supported by the Grant-in-Aid for Scientific Research from MEXT of Japan [Grants No. 17K05517, No. 25220711, and No. 15H05852 (KAKENHI on Innovative Areas “Topological Materials Science”)]. This work is supported by NSERC, CIfAR and Max Planck - UBC Centre for Quantum Materials.\end{acknowledgments}

\bibliography{a3.bib}

\clearpage
\appendix
\section{Low-energy Weyl Hamiltonian}
\label{Low-energy Weyl Hamiltonian}
    In the main text Section~\ref{Strain induced gauge field}, we find that strain generally shifts the Weyl points and effectively works as a pseudo magnetic gauge potential. In this section, we will examine this from another point of view. 
    
As the gauge field effect occurs in the low-energy sector, it will be sufficient to consider $\tilde{\mathcal{H}}_{\bm k_W+\bm q}$ defined in Eq.\ (\ref{strain_H}), namely
\begin{equation}
\tilde{\mathcal{H}}_{\bm k_W+\bm q} = \mathcal{H}_{\bm k_W} + h_{\bm q} + \delta \mathcal{H}_{\bm k_W}
\end{equation}

\noindent where $\mathcal{H}_{\bm k_W}$ is the Hamiltonian at Weyl points $\bm k_W = (0,0,\eta Q)$, $h_{\bm q}$ is the linearized Hamiltonian, and $\delta \mathcal{H}_{\bm k_W}$ is induced by strain. To keep our derivation transparent, we perform the following substitution
\begin{gather} \label{b2}
\tilde{x} = \frac{\hbar v_F}{a} (q_xa + \eta w_{31}\sin Qa) 
\\
\tilde{y} = \frac{\hbar v_F}{a} (q_ya + \eta w_{32}\sin Qa)
\\
\tilde{z}_1 = -\eta t_d \sin Qa \Big( q_za + \eta w_{33}\frac{t_s + t_d\cos Qa}{t_d\sin Qa} \Big)
\\
\tilde{z}_2 = t_d \cos Qa (q_za - \eta w_{33} \tan Qa)
\\
z_1 = t_s +t_d \cos k_za
\\
z_2 = t_d \sin k_za \label{b7}
\end{gather}

\noindent With such definitions, we can write down the eigenvalues of $\mathcal{H}_{\bm k_W}$ as $\{2m, 2m, 0,0,0,0,-2m,-2m\}$. For real order parameter $\Delta \in \mathbb{R}$,  the four eigenvectors corresponding to the Weyl points are
\begin{align}
\ket{\phi_1}&=\frac{1}{\sqrt{2}}\Big(\frac{z_1-iz_2}{m},0,-1,0,0,0,0,\frac{\Delta}{m}\Big)^T,
\\
\ket{\phi_2}&=\frac{1}{\sqrt{2}}\Big(-\frac{\Delta}{m},0,0,0,0,1,\frac{z_1+iz_2}{m}\Big)^T,
\\
\ket{\phi_3}&=\frac{1}{\sqrt{2}}\Big(0, \frac{-z_1+iz_2}{m},0,-1,0,0,\frac{\Delta}{m},0\Big)^T,
\\
\ket{\phi_4}&=\frac{1}{\sqrt{2}}\Big(0,-\frac{\Delta}{m},0,0,1,0, \frac{-z_1-iz_2}{m},0\Big)^T.
\end{align}

\noindent We then project 
$$
h_{\bm q} + \delta \mathcal{H}_{\bm k_W} = -\tilde{x} s^y\sigma^z\tau^z + \tilde{y} s^x\sigma^z + \tilde{z}_1 \sigma^x\tau^z + \tilde{z}_2 \sigma^y\tau^z
$$
onto the four-dimensional Hilbert space spanned by $\ket{\phi_{i=1,2,3,4}}$. We get
\begin{multline}
(h_{\bm q} + \delta \mathcal{H}_{\bm k_W}) _\phi = 
\\
\begin{pmatrix}
-\frac{z_1\tilde{z}_1+z_2\tilde{z}_2}{m}&&0&&
-i\tilde{x}-\tilde{y}&&0
\\
0&&-\frac{z_1\tilde{z}_1+z_2\tilde{z}_2}{m}&&0&&i\tilde{x}+\tilde{y}
\\
i\tilde{x}-\tilde{y}&&0&&\frac{z_1\tilde{z}_1+z_2\tilde{z}_2}{m}&&0
\\
0&&-i\tilde{x}+\tilde{y}&&0&&\frac{z_1\tilde{z}_1+z_2\tilde{z}_2}{m}
\end{pmatrix}
\end{multline}
The projected $4\times 4$ matrix Hamiltonian can be written in terms of standard Dirac matrices which we express as a tensor product of  Pauli matrices $\alpha$ and $\beta$ as
\begin{multline}
(h_{\bm q} + \delta \mathcal{H}_{\bm k_W})_\phi = \frac{\hbar v_F}{a} (q_xa + \eta w_{31}\sin Qa) \alpha^z\beta^y
\\
-\frac{\hbar v_F}{a} (q_ya + \eta w_{32}\sin Qa) \alpha^z\beta^x
\\
\frac{\eta t_s t_d \sin aQ}{m} \bigg(q_za + \eta w_{33} \frac{m^2-\Delta^2}{t_s t_d \sin aQ}\bigg) \beta^z
\end{multline}
From here, one can read off  the strain-induced a gauge field
\begin{equation} 
\mathcal{A} = - \frac{ \eta \hbar}{ea} \Big(w_{31} \sin Qa , w_{32} \sin Qa , w_{33} \frac{m^2-\Delta^2}{t_st_d \sin Qa}\Big)
\end{equation}
consistent with Eq.\ (\ref{A}) in the main text.

\section{Weyl superconductor with chemical potential}
\label{Weyl superconductor with chemical potential}
In the main text, we studied a Weyl superconductor multilayer with the chemical potential of the TI set to zero. In this case, the Fermi surface of the Weyl SC shrinks to two points (Weyl nodes). To observe quantum oscillation, we need a finite size Fermi surface, and this can be achieved by switching on the chemical potential of the TI layers. In this section, we will show that with nonzero chemical potential, implemented in  Eq.\ (\ref{Hk}) by adding an extra term $-\mu \tau^z$, the 2-fold degeneracy of the spectrum Eq.\ (\ref{spectrum}) will lift. One copy of the quasiparticle spectrum will move up by $\mu$ and the other  will move down by $\mu$ creating a Fermi surface with a nonzero volume.

To see this, we write down the following Hamiltonian
\begin{equation}
\mathcal{H}_{\bm k} - \mu \tau^z = H_0 + V
\end{equation}
where
\begin{gather}
H_0 = m_0 s^z \tau^z + z_1 \sigma^x \tau^z + z_2 \sigma^y \tau^z -\Delta s^y \tau^y
\\
V =-\mu \tau^z + y s^x\sigma^z - xs^y \sigma^z \tau^z 
\end{gather}
 Again, for simplicity we assume real $\Delta$ and define
\begin{gather}
m_0 = m - 4m' +2m' \cos k_xa + 2m' \cos k_ya
\\
x = \frac{\hbar v_F}{a} \sin k_{x}
\\
y = \frac{\hbar v_F}{a} \sin k_{y}
\end{gather}
Based on the parameter values we have chosen in the main text, $\Delta$ is of the same order as $z_2$, but is one order of magnitude smaller than $z_1$ and $m_0$. We set chemical potential $\mu$ to be an order of magnitude smaller than $\Delta$. This allows us to treat $V$ as a perturbation to $H_0$, whose low-energy eigenvectors can be easily resolved as
\begin{widetext}
\begin{align}
\ket{\phi_{D_1,1}} & =\frac{1}{\sqrt{2}} \Big(\frac{z_1- i z_2}{\sqrt{z_1^2+z_2^2 + \Delta^2}},0,-1,0,0,0,0,\frac{\Delta}{\sqrt{z_1^2+z_2^2 + \Delta^2}}\Big)^T,
\\
\ket{\phi_{D_1,2}} & =\frac{1}{\sqrt{2}} \Big(\frac{-\Delta}{\sqrt{z_1^2+z_2^2 + \Delta^2}},0,0,0,0,1,0,\frac{z_1+iz_2}{\sqrt{z_1^2+z_2^2 + \Delta^2}}\Big)^T,
\\
\ket{\phi_{D_2,1}} & =\frac{1}{\sqrt{2}} \Big(0,\frac{-z_1+ i z_2}{\sqrt{z_1^2+z_2^2 + \Delta^2}},0,-1,0,0,\frac{\Delta}{\sqrt{z_1^2+z_2^2 + \Delta^2}},0\Big)^T,
\\
\ket{\phi_{D_2,2}} & =\frac{1}{\sqrt{2}} \Big(0,\frac{-\Delta}{\sqrt{z_1^2+z_2^2 + \Delta^2}},0,0,1,0,\frac{-z_1-iz_2}{\sqrt{z_1^2+z_2^2 + \Delta^2}},0\Big)^T.
\end{align}
\end{widetext}
These correspond to the  degenerate subspace $D_1$ with eigenvalue 
\begin{equation}
E_{D_1,1(2)}^{(0)} = m_0-\sqrt{z_1^2+z_2^2 + \Delta^2}
\end{equation}
and  $D_2$ with eigenvalue 
\begin{equation}
E_{D_2,1(2)}^{(0)} = \sqrt{z_1^2+z_2^2 + \Delta^2}-m_0
\end{equation}
We then project the perturbation $V$ to the degenerate subspaces $D_1$ and $D_2$, respectively. In a compact form, it reads
\begin{equation}
V_{D_i} = 
\begin{pmatrix}
\braket{\phi_{D_i,1} | V | \phi_{D_i,1}} && \braket{\phi_{D_i,1} | V | \phi_{D_i,2}} \\
\braket{\phi_{D_i,2} | V | \phi_{D_i,1}} && \braket{\phi_{D_i,2} | V | \phi_{D_i,2}}
\end{pmatrix} 
\end{equation}
where $i=1,2$. Individually, we can write $V_{D_i}$ in terms of the Pauli matrix $\bm \nu$
\begin{equation}
\begin{array}{l}
V_{D_1} = d_x \nu^x + d_y \nu^y + d_z \nu^z 
\\
V_{D_2} = -d_x \nu^x - d_y \nu^y + d_z \nu^z 
\end{array}
\end{equation}
where
\begin{align}
d_x &= \frac{z_1 \Delta}{z_1^2+z_2^2+\Delta^2} \mu
\\
d_y &= - \frac{z_2 \Delta}{z_1^2+z_2^2+\Delta^2} \mu
\\
d_z &= - \frac{z_1^2 + z_2^2}{z_1^2+z_2^2+\Delta^2} \mu
\end{align}

Matrices $V_{D_1}$ and $V_{D_2}$ can be diagonalized through unitary transformations
\begin{equation}
U_{D_i}^{-1} V_{D_i} U_{D_i} = \text{diag}(d,-d) \qquad i=1,2
\end{equation}

\noindent where
\begin{equation}
d = \sqrt{\frac{z_1^2+z_2^2}{z_1^2+z_2^2+\Delta^2}} \mu \approx \mu
\end{equation}

\noindent because for our purpose $\Delta^2 \ll z_1^2+z_2^2$. The transformation matrices are
\begin{align}
U_{D_1} &=
\begin{pmatrix}
\sqrt{\frac{d+d_z}{2d}}\frac{d_x - id_y}{\sqrt{d_x^2+d_y^2}}
&&
-\sqrt{\frac{d-d_z}{2d}}
\\
\sqrt{\frac{d-d_z}{2d}}
&&
\sqrt{\frac{d+d_z}{2d}} \frac{d_x + id_y}{\sqrt{d_x^2+d_y^2}}
\end{pmatrix},
\\
U_{D_2} &=
\begin{pmatrix}
\sqrt{\frac{d+d_z}{2d}}\frac{-d_x + id_y}{\sqrt{d_x^2+d_y^2}}
&&
-\sqrt{\frac{d-d_z}{2d}}
\\
\sqrt{\frac{d-d_z}{2d}}
&&
\sqrt{\frac{d+d_z}{2d}} \frac{-d_x - id_y}{\sqrt{d_x^2+d_y^2}}
\end{pmatrix}.
\end{align}
We can immediately write down the first order correction to energy
\begin{equation}
E_{D_1,1(2)}^{(1)} = E_{D_2,1(2)}^{(1)} =  \pm d \approx \pm \mu.
\end{equation}
Because the 2-fold degeneracy is lifted, the zeroth order eigenvectors are now uniquely determined 
\begin{gather}
(\ket{\tilde \phi_{D_1,1}},  \ket{\tilde \phi_{D_1,2}})= (\ket{\phi_{D_1,1}},  \ket{\phi_{D_1,2}}) U_{D_1},
\\
(\ket{\tilde \phi_{D_2,1}},  \ket{\tilde \phi_{D_2,2}})= (\ket{\phi_{D_2,1}},  \ket{\phi_{D_2,2}}) U_{D_2}.
\end{gather}
We may now calculate the second order correction to the  energy
\begin{multline}
E_{D_1,1(2)}^{(2)} =\sum\limits_{\alpha \in D_2 } \frac{|\braket{\tilde \phi_\alpha | V | \tilde \phi_{D_1,1(2)}}|^2}{E_{D_1}-E_{\alpha}}
\\
= \frac{x^2+y^2}{2(m_0-\sqrt{z_1^2+z_2^2+\Delta^2})}
\end{multline}

\begin{multline}
E_{D_2,1(2)}^{(2)} =\sum\limits_{\alpha \in D_1 } \frac{|\braket{\tilde \phi_\alpha | V | \tilde \phi_{D_2,1(2)}}|^2}{E_{D_2}-E_{\alpha}}
\\
= - \frac{x^2+y^2}{2(m_0-\sqrt{z_1^2+z_2^2+\Delta^2})}
\end{multline}
where we ignore the contribution from high-energy sector, if any. This is because for high energies $E_\alpha = \pm (m_0 + \sqrt{z_1^2+z_2^2+\Delta^2})$ the denominator in the second order correction is either $\pm 2m_0$ or $\pm 2\sqrt{z_1^2+z_2^2+\Delta^2}$, whose magnitude is much larger than that of $\pm 2(m_0-\sqrt{z_1^2+z_2^2+\Delta^2})$ and thus are less important. Combining all the corrections, we can estimate the quasiparticle energy at nozero $\mu$ as
\begin{align}
E_{D_1,1(2)} & \approx \sqrt{x^2+y^2+(m_0-\sqrt{z_1^2+z_2^2+\Delta^2})^2} \pm \mu,
\\
E_{D_2,1(2)} & \approx -\sqrt{x^2+y^2+(m_0-\sqrt{z_1^2+z_2^2+\Delta^2})^2} \pm \mu.
\end{align}

We observe that to leading order the spectrum of Weyl superconductor multilayer is now biased. The original 2-fold degeneracy in Eq.\ (\ref{spectrum}) has been lifted. One copy of spectrum moves up while the other copy moves down. As a result, the strain induced pseudo Landau levels will also be biased as in Eq.\ (\ref{biased Dirac-Landau})

\section{Thermal conductivity}
\label{Thermal conductivity}
In this section, we will derive the expression for the thermal conductivity given  in Eq.\ (\ref{Wiedemann-Franz}) of main text. Our starting point is Eq.\ (\ref{kappaxx1}).
In order to calculate $\bm \kappa$ analytically, we introduce the auxiliary tensor
\begin{equation}
\bm \sigma (\epsilon) = \sum\limits_n\sum\limits_{\bm k} \tau_n(\bm k) \delta(\epsilon-E_n(\bm k)) \bm v_n(\bm k) \bm v_n(\bm k),
\end{equation}
which may be understood as a thermal analogue of the usual conductivity tensor. It is easy to see that
\begin{equation}
\bm \kappa = \frac{1}{T} \int\nolimits_{-\infty}^{+\infty} d\epsilon \epsilon^2 \bm \sigma(\epsilon) \bigg(- \frac{\partial f}{\partial \epsilon} \bigg)
\end{equation}
We further define an auxiliary function
\begin{equation}
K(\epsilon)= \epsilon^2 \bm \sigma(\epsilon)
\end{equation}
through which  the thermal conductivity can be written as
\begin{equation}
\bm \kappa = \frac{1}{T} \int\nolimits_{-\infty}^{+\infty} d\epsilon \sum\limits_{s=1}^{\infty} \frac{1}{s!} \frac{d^s K}{d\epsilon^s}\bigg|_{0} \epsilon^s \bigg(- \frac{\partial f}{\partial \epsilon} \bigg).
\end{equation}
 Note that $\frac{\partial f}{\partial \epsilon}$ is an even function of $\epsilon$. Therefore, we only need to consider even $s$. The thermal conductivity is further simplified as
\begin{equation}
\bm \kappa = \frac{1}{T} \sum\limits_{s=1}^{\infty} \int\nolimits_{-\infty}^{+\infty} d\epsilon  \frac{(k_BT)^{2s}}{(2s)!} \bigg(\frac{\epsilon}{k_BT}\bigg)^{2s} \bigg(- \frac{\partial f}{\partial \epsilon} \bigg) \frac{d^{2s} K}{d\epsilon^{2s}}\bigg|_{0} 
\end{equation}
Define $x=\frac{\epsilon}{k_BT}$ and
\begin{multline}
a_s=\int\nolimits_{-\infty}^{+\infty} dx  \frac{x^{2s}}{(2s)!} \bigg( -\frac{d}{dx} \frac{1}{e^x+1}\bigg) 
\\
= \frac{2}{\Gamma(2s)}\int\nolimits_{0}^{+\infty} dx\frac{x^{2s-1}}{e^x+1}
\\
=2\eta(2s)=2(1-2^{1-2s})\zeta(2s)
\end{multline}
where $\Gamma(s)=\int\nolimits_0^{+\infty} dx\frac{x^{s-1}}{e^x}$, $\eta(s)=\int\nolimits_0^{+\infty} dx\frac{x^{s-1}}{e^x+1}$, and $\zeta(s)=\int\nolimits_0^{+\infty} dx\frac{x^{s-1}}{e^x-1}$ are Gamma function, Dirichlet eta function, and Riemann zeta function, respectively. The thermal conductivity now reads 
\begin{equation}
\bm \kappa = \frac{1}{T} \sum\limits_{s=1}^{\infty} 2(1-2^{1-2s})\zeta(2s)
 \frac{d^{2s} K}{d\epsilon^{2s}}\bigg|_{0} (k_BT)^{2s}
\end{equation}
For low temperatures  $k_BT\ll\mu$, we keep only the $s=1$ term and use $\zeta(2)=\frac{\pi^2}{6}$ to get
\begin{multline}
\bm \kappa = \frac{1}{T}\frac{\pi^2k_B^2T^2}{3} \bm \sigma(0) =
\\
 \frac{\pi^2k_B^2T}{3}\sum\limits_n\sum\limits_{\bm k} \tau_n(\bm k) \delta(E_n(\bm k)) \bm v_n(\bm k) \bm v_n(\bm k)
\end{multline}
This relation can be regarded as the Wiedemann-Franz law for Bogoliubov quasiparticles. 

The scattering time can be determined by Fermi's golden rule
\begin{multline}
\frac{1}{\tau_n(\bm k)}=\frac{2\pi}{\hbar}\sum\limits_{n'} \sum\limits_{\bm k'} 
 |\braket{n'\bm k'|V(\bm r)_{\text{imp}}\tau^z|n\bm k}|^2
 \\
 \times  \delta(E_n(\bm k)-E_{n'}(\bm k'))
\end{multline}
 where $\ket{n \bm k}$ is the eigenvector of chemical potential biased Weyl superconductor under strain, characterized by the Hamiltonian $H-\mu \tau^z$. As discussed in Appendix~\ref{Weyl superconductor with chemical potential}, when $\mu^2 \ll \Delta^2 \ll t_s^2+t_d^2+2t_st_d\cos k_za$, we can use perturbative calculation to write down the Schr\"odinger equations for Weyl superconductor with TI layer chemical potential $\mu \neq 0$ and $\mu=0$, respectively.
\begin{align}
(H-\mu \tau^z) \ket{n \bm k^{0}} & \approx \tilde{\epsilon}_n(\bm k) \ket{n \bm k^{0}} \pm \mu \ket{n \bm k^{0}}, \label{d11}
\\
H\ket{n \bm k^{0}} &= \tilde{\epsilon}_n(\bm k) \ket{n \bm k^{0}}, \label{d12}
\end{align}
where $\tilde{\epsilon}_n(\bm k)$ is determined by Eq.\ (\ref{En}) and $\ket{n \bm k^{0}}$ is the exact eigenvector of $H$ and the zeroth order eigenvector of $H-\mu \tau^z$. If apply $\bra{n' \bm k'^{0}}$ to Eq.\ (\ref{d11}) and Eq.\ (\ref{d12}) and subtract, we get
\begin{equation}
\braket{n' \bm k'^{0}| \tau^z |n \bm k^{0}} \approx \pm \braket{n' \bm k'^{0} | n \bm k^{0}}
\end{equation} 
then we can approximate $\tau_n(\bm k)$ by
\begin{multline}
\frac{1}{\tau_n(\bm k)} \approx \frac{2\pi}{\hbar}\sum\limits_{n'} \sum\limits_{\bm k'}  |\braket{n'\bm k'^0|V(\bm r)_{\text{imp}}| n\bm k^0}|^2
 \\
 \times  \delta(E_n(\bm k)-E_{n'}(\bm k'))
\end{multline}
The righthand side is the same as the scattering rate of a Weyl semimetal \cite{liu2017} with electronic structure characterized by $E_n(\bm k)$. Therefore, the scattering rate in a Weyl superconductor should also be the same which leads to Eq.\ (\ref{tau}) in the main text.

\section{Vortex lattice}
\label{Vortex lattice}
In this section we study Weyl superconductors under real magnetic field $\bm B$ and compare the results to Section \ref{Strain induced gauge field} in the main text. Due to the Meissner effect, $\bm B$ field is known to generate quasiparticle Bloch waves rather than Dirac-Landau levels in two-dimensional nodal SC, such as those with a $d$-wave symmetry of the gap function \cite{franz2000, vafek2001}. It is however not known how this result translates to three-dimensional Weyl SC. 

We consider magnetic field along $z$-direction, so that $k_z$ remains a good quantum number. Thus, the system can, in principle, stay gapless. To study the vortex lattice, we write Eq.\ (\ref{H}) as
\begin{equation}
H=\frac{1}{2} \sum\limits_{\bm k} \Psi_{\bm r}^\dagger \mathcal{H}_{\bm r} \Psi_{\bm r} = \frac{1}{2} \sum\limits_{\bm k} \Psi_{\bm r}^\dagger \begin{pmatrix} \mathcal{H}_{\bm r}^{11} &&\mathcal{H}_{\bm r}^{12}\\ \mathcal{H}_{\bm r}^{21} && \mathcal{H}_{\bm r}^{22} \end{pmatrix} \Psi_{\bm r}
\end{equation}
with the real space basis to be written as $\Psi_{\bm r} = (c_{\bm r, \uparrow, 1}, c_{\bm r, \downarrow, 1}, c_{\bm r, \uparrow, 2}, c_{\bm r, \downarrow, 2}, c_{\bm r, \uparrow, 1}^\dagger, c_{\bm r, \downarrow, 1}^\dagger, c_{\bm r, \uparrow, 2}^\dagger, c_{\bm r, \downarrow, 2}^\dagger)^T$ and the blocks are defined as
\begin{widetext}
\begin{equation}
\mathcal{H}_{\bm r}^{11} = \\
\begin{pmatrix}
\begin{smallmatrix}
m-4b+b\sum\limits_{\bm \delta} \hat{s}_{\bm \delta} &&
-i\frac{\hbar v_F}{2a}\sum\limits_{\bm \delta} \hat{\eta}_{\bm \delta}^* &&
t_s+t_de^{-ik_za}&&
0\\
-i\frac{\hbar v_F}{2a}\sum\limits_{\bm \delta} \hat{\eta}_{\bm \delta} &&
-m+4b-b\sum\limits_{\bm \delta} \hat{s}_{\bm \delta} &&
0&&
t_s+t_de^{-ik_za}\\
t_s+t_de^{ik_za}&&
0&&
m-4b+b\sum\limits_{\bm \delta} \hat{s}_{\bm \delta} &&
i\frac{\hbar v_F}{2a}\sum\limits_{\bm \delta} \hat{\eta}_{\bm \delta}^* \\
0&&
t_s+t_de^{ik_za}&&
i\frac{\hbar v_F}{2a}\sum\limits_{\bm \delta} \hat{\eta}_{\bm \delta} &&
-m+4b-b\sum\limits_{\bm \delta} \hat{s}_{\bm \delta}
\end{smallmatrix}
\end{pmatrix}
\end{equation}
\begin{equation}
\mathcal{H}_{\bm r}^{22} = \\
\begin{pmatrix}
\begin{smallmatrix}
-m+4b-b\sum\limits_{\bm \delta} \hat{s}_{\bm \delta} &&
-i\frac{\hbar v_F}{2a}\sum\limits_{\bm \delta} \hat{\eta}_{\bm \delta} &&
-t_s-t_de^{-ik_za}&&
0\\
-i\frac{\hbar v_F}{2a}\sum\limits_{\bm \delta} \hat{\eta}_{\bm \delta}^* &&
m-4b+b\sum\limits_{\bm \delta} \hat{s}_{\bm \delta} &&
0&&
-t_s-t_de^{-ik_za}\\
-t_s-t_de^{ik_za}&&
0&&
-m+4b-b\sum\limits_{\bm \delta} \hat{s}_{\bm \delta} &&
i\frac{\hbar v_F}{2a}\sum\limits_{\bm \delta} \hat{\eta}_{\bm \delta} \\
0&&
-t_s-t_de^{ik_za}&&
i\frac{\hbar v_F}{2a}\sum\limits_{\bm \delta} \hat{\eta}_{\bm \delta}^* &&
m-4b+b\sum\limits_{\bm \delta} \hat{s}_{\bm \delta}
\end{smallmatrix}
\end{pmatrix}
\end{equation}
\begin{equation}
\mathcal{H}_{\bm r}^{12} =
\begin{pmatrix}
\begin{smallmatrix}
0&&\Delta&&0&&0\\
-\Delta&&0&&0&&0\\
0&&0&&0&&\Delta\\
0&&0&&-\Delta&&0
\end{smallmatrix}
\end{pmatrix}
\qquad 
\mathcal{H}_{\bm r}^{21}=
\begin{pmatrix}
\begin{smallmatrix}
0&&-\Delta^*&&0&&0\\
\Delta^*&&0&&0&&0\\
0&&0&&0&&-\Delta^*\\
0&&0&&\Delta^*&&0
\end{smallmatrix}
\end{pmatrix}
\end{equation}
\end{widetext}
Here the shift operator is defined as
\begin{equation}
\hat{s}_{\bm \delta} f(\bm r) = f(\bm r + \bm \delta) \qquad \bm \delta = \pm a \hat x, \pm a \hat y
\end{equation}
and
\begin{equation}
	\hat{\eta}_{\bm{\delta}}=
	\begin{cases}
	\mp i \hat{s}_{\bm{\delta}} &\mbox{ if \(\bm{\delta}=\pm a \hat{x}\)}\\
	\pm  \hat{s}_{\bm{\delta}} &\mbox{ if \(\bm{\delta}=\pm a \hat{y}\)}
	\end{cases}.
\end{equation}

\begin{figure}[t]
\includegraphics[width=5cm]{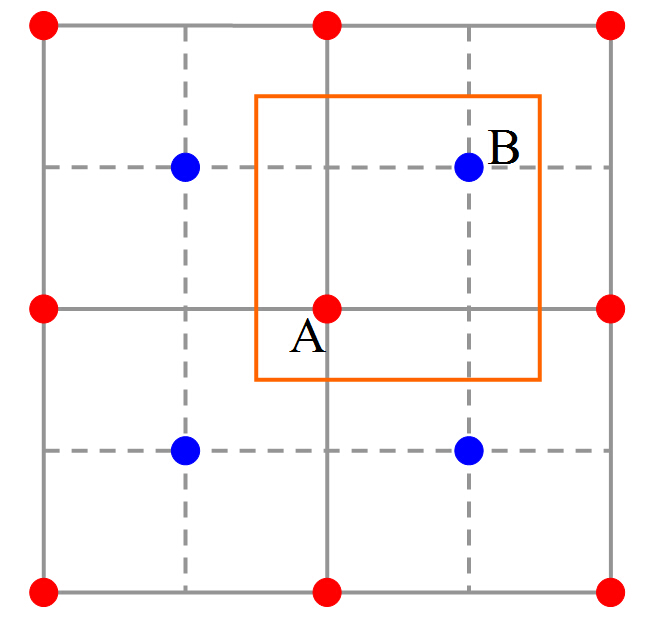} 
\caption{Schematic plot of square vortex lattice. The red and blue dots correspond to two vortex sublattices. The orange square is the magnetic unit cell with vortices placed on the diagonal. The dimension of the magnetic unit cell is chosen to be $L=30a$ in the simulation.}
\label{fig8}
\end{figure}

 In order to model vortex lattice, the phase of $\Delta({\bm r})=\Delta_0e^{i \phi(\bm r)}$ is taken to wind by $2\pi$ around each vortex center. We solve the problem 
by performing a unitary transformation \cite{franz2000} in the Nambu space defined by 
\begin{equation}
U=
\begin{pmatrix}
e^{i \phi_A(\bm r)} && 0\\
0 && e^{- i \phi_B(\bm r)}
\end{pmatrix},
\end{equation}
where we have partitioned vortices into two sublattices A and B such that $\phi_A(\bm r)+\phi_B(\bm r)=\phi(\bm r)$. This removes the phase winding from the off-diagonal part of the Hamiltonian and makes it periodic in real space with a unit cell depited in Fig.\ \ref{fig8}.  

The eigenstates of the transformed Hamiltonian are Bloch waves \cite{franz2000,vafek2001,liu2015} that read $\Phi_{n \bm K}(\bm r) = e^{i \bm K \cdot \bm r} [U_{n \bm K}(\bm r), V_{n \bm K}(\bm r)]^T$ with crystal momentum $\bm K$ associated with the vortex lattice (Fig.~\ref{fig8}). The BdG type Bloch Hamiltonian is $H_{\bm K}= e^{-i \bm K \cdot \bm r} U^{-1} H_{\bm r} U e^{i \bm K \cdot \bm r}$ with its 4 blocks $H_{\bm K}^{ij}= e^{-i \bm K \cdot \bm r} U^{-1} H_{\bm r}^{ij} U e^{i \bm K \cdot \bm r}$ defined as \begin{widetext}
\begin{equation}
\mathcal{H}_{\bm K}^{11} = \\
\begin{pmatrix}
\begin{smallmatrix}
m-4b+b\sum\limits_{\bm \delta} e^{i \bm K \cdot \bm \delta} e^{i \mathcal{V}_{\bm \delta}^A} \hat{s}_{\bm \delta} &&
-i\frac{\hbar v_F}{2a}\sum\limits_{\bm \delta} e^{i \bm K \cdot \bm \delta} e^{i \mathcal{V}_{\bm \delta}^A} \hat{\eta}_{\bm \delta}^* &&
t_s+t_de^{-ik_za}&&
0\\
-i\frac{\hbar v_F}{2a}\sum\limits_{\bm \delta} e^{i \bm K \cdot \bm \delta} e^{i \mathcal{V}_{\bm \delta}^A}\hat{\eta}_{\bm \delta} &&
-m+4b-b\sum\limits_{\bm \delta}e^{i \bm K \cdot \bm \delta} e^{i \mathcal{V}_{\bm \delta}^A} \hat{s}_{\bm \delta} &&
0&&
t_s+t_de^{-ik_za}\\
t_s+t_de^{ik_za}&&
0&&
m-4b+b\sum\limits_{\bm \delta}e^{i \bm K \cdot \bm \delta} e^{i \mathcal{V}_{\bm \delta}^A} \hat{s}_{\bm \delta} &&
i\frac{\hbar v_F}{2a}\sum\limits_{\bm \delta}e^{i \bm K \cdot \bm \delta} e^{i \mathcal{V}_{\bm \delta}^A} \hat{\eta}_{\bm \delta}^* \\
0&&
t_s+t_de^{ik_za}&&
i\frac{\hbar v_F}{2a}\sum\limits_{\bm \delta}e^{i \bm K \cdot \bm \delta} e^{i \mathcal{V}_{\bm \delta}^A} \hat{\eta}_{\bm \delta} &&
-m+4b-b\sum\limits_{\bm \delta}e^{i \bm K \cdot \bm \delta} e^{i \mathcal{V}_{\bm \delta}^A} \hat{s}_{\bm \delta}
\end{smallmatrix}
\end{pmatrix}
\end{equation}

\begin{equation}
\mathcal{H}_{\bm K}^{22} =
\begin{pmatrix}
\begin{smallmatrix}
-m+4b-b\sum\limits_{\bm \delta} e^{i \bm K \cdot \bm \delta} e^{- i \mathcal{V}_{\bm \delta}^B} \hat{s}_{\bm \delta} &&
-i\frac{\hbar v_F}{2a}\sum\limits_{\bm \delta} e^{i \bm K \cdot \bm \delta} e^{- i \mathcal{V}_{\bm \delta}^B}\hat{\eta}_{\bm \delta} &&
-t_s-t_de^{-ik_za}&&
0\\
-i\frac{\hbar v_F}{2a}\sum\limits_{\bm \delta} e^{i \bm K \cdot \bm \delta} e^{- i \mathcal{V}_{\bm \delta}^B}\hat{\eta}_{\bm \delta}^* &&
m-4b+b\sum\limits_{\bm \delta}e^{i \bm K \cdot \bm \delta} e^{- i \mathcal{V}_{\bm \delta}^B} \hat{s}_{\bm \delta} &&
0&&
-t_s-t_de^{-ik_za}\\
-t_s-t_de^{ik_za}&&
0&&
-m+4b-b\sum\limits_{\bm \delta} e^{i \bm K \cdot \bm \delta} e^{- i \mathcal{V}_{\bm \delta}^B}\hat{s}_{\bm \delta} &&
i\frac{\hbar v_F}{2a}\sum\limits_{\bm \delta}e^{i \bm K \cdot \bm \delta} e^{- i \mathcal{V}_{\bm \delta}^B} \hat{\eta}_{\bm \delta} \\
0&&
-t_s-t_de^{ik_za}&&
i\frac{\hbar v_F}{2a}\sum\limits_{\bm \delta}e^{i \bm K \cdot \bm \delta} e^{- i \mathcal{V}_{\bm \delta}^B} \hat{\eta}_{\bm \delta}^* &&
m-4b+b\sum\limits_{\bm \delta} e^{i \bm K \cdot \bm \delta} e^{- i \mathcal{V}_{\bm \delta}^B}\hat{s}_{\bm \delta}
\end{smallmatrix}
\end{pmatrix}
\end{equation}

\begin{equation}
\mathcal{H}_{\bm K}^{12} =
\begin{pmatrix}
\begin{smallmatrix}
0&&\Delta&&0&&0\\
-\Delta&&0&&0&&0\\
0&&0&&0&&\Delta\\
0&&0&&-\Delta&&0
\end{smallmatrix}
\end{pmatrix}
\qquad
\mathcal{H}_{\bm K}^{21}=
\begin{pmatrix}
\begin{smallmatrix}
0&&-\Delta^*&&0&&0\\
\Delta^*&&0&&0&&0\\
0&&0&&0&&-\Delta^*\\
0&&0&&\Delta^*&&0
\end{smallmatrix}
\end{pmatrix}
\end{equation}
\end{widetext}
 where the phase factors associated with two types of vortices are
\begin{equation}\label{Vs}
\mathcal{V}_{\bm \delta}^{\mu} =\frac{m}{\hbar} \int\nolimits_{\bm r}^{\bm r + \bm \delta} \bm v_{s}^{\mu}(\bm r) \cdot d \bm l \qquad \mu=A,B.
\end{equation}
The integral is along the bond connecting lattice point $\bm r$ to its nearest neighbor $\bm r + \bm \delta$. The superfluid velocity is
\begin{equation} \label{vs}
\bm v_{s}^{\mu}(\bm r) =\frac{\hbar}{m} \big( \nabla \phi^\mu - \frac{e}{\hbar} \bm A(\bm r) \big) \qquad\mu=A,B.
\end{equation}

\begin{figure*}[ht]
\includegraphics[width=18cm]{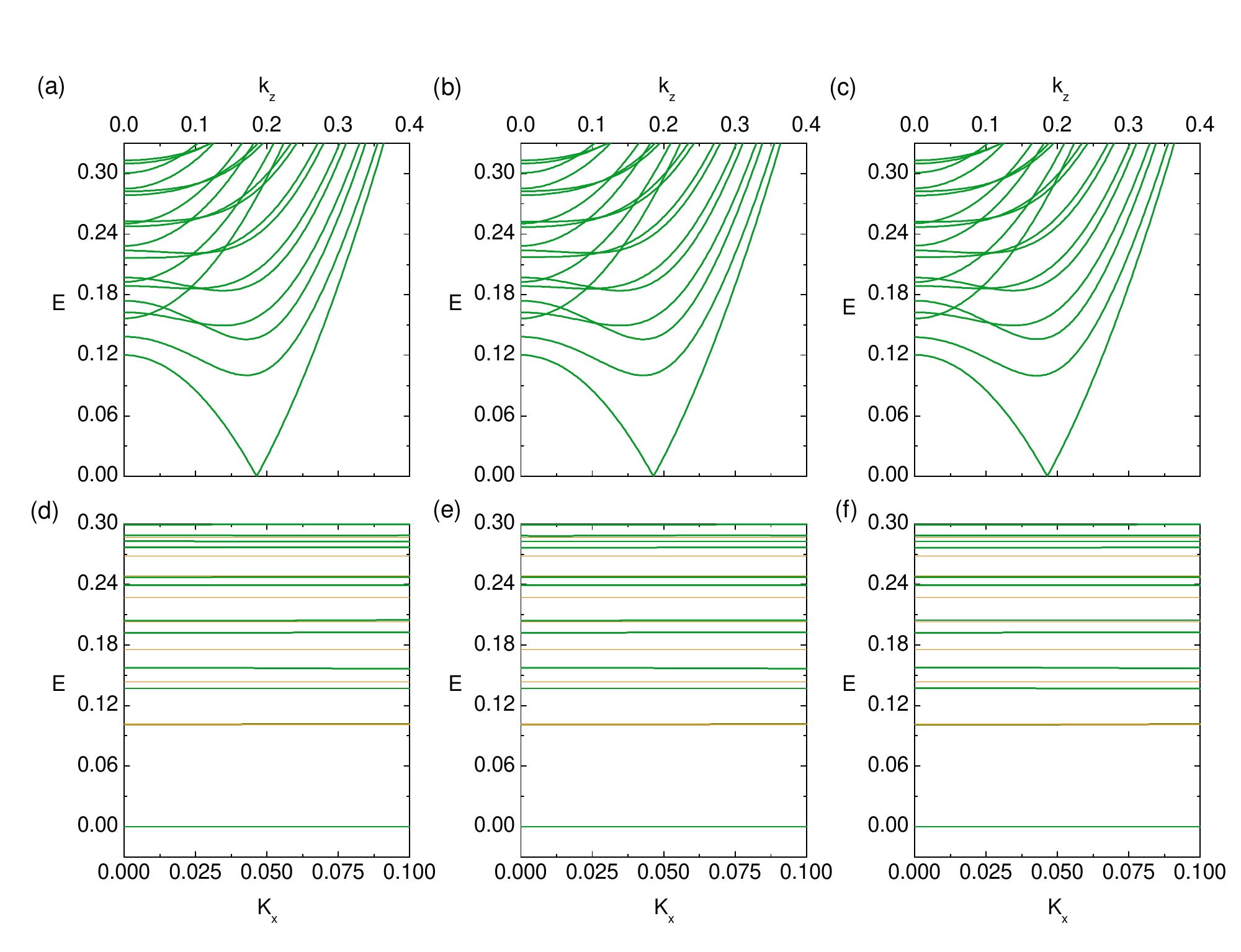}
\caption{Spectra of Weyl superconductor with vortex lattice. The size of magnetic unit cell is $L\times L = 30a \times 30a$. The spacings between two vortices in the magnetic unit cell are (a) $\bm d = (15a, 15a)$ (b) $\bm d = (10a,10a)$ (c) $\bm d = (5a,5a)$ (d) $\bm d = (15a, 15a)$ (e) $\bm d = (10a, 10a)$ (f) $\bm d = (5a, 5a)$ The orange curves in panel (d)-(f) are analytical Dirac-Landau levels with $n=1$ band matched to the numerics. }
\label{fig9}
\end{figure*}

Following the standard derivation \cite{vafek2001}  an expression for $\mathcal{V}_{\bm \delta}^{\mu}$ can be derived in terms of summation over the reciprocal lattice vectors $\bm G$ of the vortex lattice, 
\begin{equation} \label{practical_V}
	\mathcal{V}^\mu_{\bm \delta}(\bm r) = \frac{2 \pi}{L^2} \sum\limits_{\bm G}\int\nolimits_{\bm r}^{\bm r + \bm \delta} e^{i \bm G \cdot (\bm r-\bm \delta^ \mu)} \frac{i \bm G \times \hat{z}}{G^2} \cdot d \bm l
\end{equation}
We apply Eq.\ (\ref{practical_V}) to the real space Hamiltonian $H_{\bm r}$ and exactly diagonalize $H_{\bm r}$ for various vortex lattice configurations. The dispersions along the $k_z$-axis are summarized in Fig.~\ref{fig9}(a-c). We observe that the Weyl points survive in as we change the A-B vortex distance within each unit cell. Surprisingly, the variation of the vortex positions barely changes the dispersion. Therefore, we conclude that the $k_z$ component of the Weyl dispersion is stable under magnetic field $B$ as long as vortices form a periodic lattice. 

Dispersion in the $K_x-K_y$ plane however changes dramatically. In Fig. ~\ref{fig9}(d-f), we plot dispersions along $K_x$ for the vortex configurations used in panels (a-c). We see that the energy bands are reorganized into almost completely flat  Dirac-Landau levels which are qualitatively similar to those reported by Ref.~\cite{pacholski1}. For comparison we also indicate the expected energies $\sim\sqrt{n}$ of  Dirac-Landau levels (orange curves) by matching to the $n=0,1$ bands. It is worth noting that the deviation of numerically calculated bands (green curves) from the ideal  $\sqrt{n}$ sequence are due to the fact that Dirac-Landau levels exist only in the low-energy regime in the vicinity of the Weyl points. For our model, Lifshitz transition occurs at $E_{\text{Lif}}=0.138$. Therefore, we do not expect a perfect match to the $\sqrt{n}$ behavior beyond the lowest few energy levels.

\end{document}